\documentclass[preprintnumbers,amsmath,11pt,amssymb,floatfix,superscriptaddress,nofootinbib]{article}

\def\mytitle#1{\setcounter{equation}{0}
\setcounter{footnote}{0}
\begin{flushleft}\Large\textbf{#1}\end{flushleft}
\vspace{0.25cm}}
\def\myname#1{\leftline{{\large #1}}\vspace{-0.13cm}}
\def\myplace#1#2{\small\begin{flushleft}\textit{#1}\\
\texttt{#2}\end{flushleft}}

\def\myclassification#1{\small\noindent
Pacs no :
       #1\vspace{0.5cm}}
\usepackage{graphicx}
\begin{document}

\mytitle{Presence of Dark Energy and Dark Matter : Does Cosmic Acceleration signifies a Weak Gravitational collapse?}

\myname{Prabir Rudra~\footnote{prudra.math@gmail.com}$*$}
\vskip0.2cm\myname{Ritabrata
Biswas~\footnote{ritabrata@phys.iisc.ernet.in, biswas.ritabrata@gmail.com}$\dag$}\vskip0.2cm \myname{Ujjal Debnath~\footnote{ujjal@iucaa.ernet.in,
ujjaldebnath@yahoo.com}$*$} \vskip0.2cm
\myplace{$*$Department of Mathematics, Bengal Engineering and
Science University, Shibpur, Howrah-711 103, India.} {}
\myplace{$\dag$ Department of Physics, Indian Institute of
Science, Bangalore, India.} {}
\myclassification{04.20~-q,~~04.40~ Dg,~~97.10.~Cv.}
\vspace{1cm}
\begin{abstract}
In this work the collapsing process of a spherically symmetric
star, made of dust cloud, in the background of dark energy is
studied for two different gravity theories separately, i.e., DGP
Brane gravity and Loop Quantum gravity. Two types of dark energy
fluids, namely, Modified Chaplygin gas and Generalised Cosmic
Chaplygin gas are considered for each model. Graphs are drawn to
characterize the nature and the probable outcome of gravitational
collapse. A comparative study is done between the collapsing
process in the two different gravity theories. It is found that in
case of dark matter, there is a great possibility of collapse and
consequent formation of Black hole. In case of dark energy
possibility of collapse is far lesser compared to the other cases,
due to the large negative pressure of dark energy component. There
is an increase in mass of the cloud in case of dark matter
collapse due to matter accumulation. The mass decreases
considerably in case of dark energy due to dark energy accretion
on the cloud. In case of collapse with a combination of dark
energy and dark matter, it is found that in the absence of
interaction there is a far better possibility of formation of
black hole in DGP brane model compared to Loop quantum cosmology
model.
\end{abstract}

\newpage
\tableofcontents

\section{Introduction}\label{chap01}
The most remarkable and significant discovery in recent past in
the field of cosmology is the discovery of the fact, that our
universe is undergoing an accelerated expansion of late
\cite{Perlmutter1,Spergel1}. This event has overturned a few
stones in the traditional theories of cosmology. Einstein's
equation needed serious revisions in order to account for the
observed cosmic acceleration.

Some cosmologists succeeded in affecting proper revisions to the
Einstein's equation by introducing changes in the right side,
i.e., introducing the concept of Dark Energy (DE
hereafter)\cite{Riess1} component with negative pressure which is
responsible for the accelerated expansion of the universe. There
have been a lot of research on DE till date. Many candidates
having the potential to play the role of DE have surfaced. Among
them the DE associated with a scalar field is often called
quintessence. In the cosmological context Chaplygin gas was first
suggested as an alternative to quintessence. The earliest form of
this was known as Pure Chaplygin gas \cite{Kamenshchik1,Gorini1}.
Gradually Generalized Chaplygin gas \cite{Gorini2,Alam1,Bento1}
was formulated and finally Modified Chaplygin Gas (MCG) came to
the foreground. The equation of state followed by MCG
\cite{Benaoum1,Debnath1} is given by
\begin{equation}\label{collapse3.1}
p=A\rho_{MCG}-\frac{B}{\rho_{MCG}^\alpha},
\end{equation}
where $p$ and $\rho$ are respectively the pressure and energy
density and $0\leq\alpha\leq1$, and $~A, B>0$. MCG appears to be
consistent with the WMAP 5-year data and henceforth support the
unified model with DE and dark matter(DM) based on generalized
Chaplygin gas \cite{Setare3,Setare4,Setare5,Barriero2,Makler1}.
MCG demonstrates an increasing $\Lambda$ behaviour for the
evolution of the universe \cite{Kamenshchik1}. Recent developments
regarding MCG can be found in \cite{Lu1,Daojun1,Jing1,Debnath2}.
Recent supernovae data also favors the two-fluid cosmological
model with Chaplygin gas and matter \cite{Panotopoulos1}.

In this paper in addition to MCG we will study the gravitational
collapse of a star in presence of one more fluid known as,
Generalized Cosmic Chaplygin Gas(GCCG). Pedro F. Gonzalez-Diaz
\cite{Gonzalez1} for the first time presented the idea of GCCG.
His main motivation was to study DE accretion onto Black Hole(BH),
and by using GCCG as DE he showed that Big Rip, i.e., singularity
at a finite time , is completely out of question, which was an
essential component of the previous models. The Equation of state
for GCCG is given by,
\begin{equation}\label{collapse3.2}
p=-\rho_{GCCG}^{-\alpha}\left[C'+\left(\rho_{GCCG}^{1+\alpha}-C'\right)^{-\omega}\right]
\end{equation}
where $C'=\frac{A'}{1+\omega}-1$. ~~Here $A'$ is a constant which
can take on both positive and negative values, and
$0>\omega>-\pounds$, $\pounds$ being a positive definite constant
which can take on values larger than unity. GCCG can explain the
evolution of universe starting from dust era to $\Lambda CDM$,
radiation era, matter dominated quintessence and lastly phantom
era \cite{Chakraborty1}.

It is a known that general relativity is incomplete as a theory of
gravity. The reason being that the gravitational collapse of
physical matter produces a BH with a singularity inside, which is
a point in space-time where the curvature and energy density
diverge. At this point all mathematical structures break down
preventing any further analysis beyond the singularity. Herein
lies the motivation for extensive research in gravitational
collapse. Till date a lot of significant work has been done in
gravitational collapse starting from the pioneering works of
Oppenheimer and Snyder \cite{Oppenheimer1}. In this paper our
motivation is to study the nature and outcome of gravitational
collapse of a star made up of DM in the background of DE (of
different forms), in two types of widely known gravity theories.

The paper is organized as follows: In section \ref{chap02}, the general
formulation of the collapsing process of a cloud is given. Section
\ref{chap03} and section \ref{chap04} deals with the details of collapsing procedure in
DGP brane model and Loop quantum cosmology(LQC) respectively.  Finally
the paper ends with some concluding remarks in section \ref{chap05}.

\section{General Formulation of the Collapsing process}\label{chap02}
The flat, homogeneous and isotropic FRW model of the universe is
described by the line element
\begin{equation}\label{collapse3.3}
ds^{2}=dt^{2}+a^{2}(t)\left[dr^{2} +r^{2}\left(d\theta^{2}
+Sin^{2}\theta d\phi^{2}\right)\right]
\end{equation}
The energy conservation equation is given by
\begin{equation}\label{collapse3.4}
\dot{\rho}_{T}+ 3 \frac{\dot{a}}{a}(\rho_{T}+p_{T})=0
\end{equation}
with $\rho_{T}=\rho_{M}+\rho_{E}$ and $p_{T}=p_{M}+p_{E}$.

The interaction $Q(t)$ between DM and DE can be expressed as
\begin{equation}\label{collapse3.5}
\dot{\rho}_{M}+3 \frac{\dot{a}}{a} \rho_{M}=Q
\end{equation}
\begin{equation}\label{collapse3.6}
\dot{\rho}_{E}+ 3 \frac{\dot{a}}{a}(\rho_{E}+p_{E})=-Q
\end{equation}
Now, if we consider gravitational collapse of a spherical cloud
consists of above DM and DE distribution and is bounded by the
surface $\Sigma : r=r_{\Sigma}$ then the metric on it can be
written as
\begin{equation}\label{collapse3.7}
ds^{2}=dT^{2}-R^{2} (T) \{d\theta^{2} +Sin^{2} \theta d\phi^{2}\}
\end{equation}
Thus on $\Sigma : T=t$ and $R(T)=r_{\Sigma} a(T)$ where $
R(r,t)\equiv r a(t)$ is the geometrical radius of the two spheres
$t,r =$ constant. Also the total mass of the collapsing cloud is
given by
\begin{equation}\label{collapse3.8}
M(T) = \left. m(r,t)\right|_{r=r_{\Sigma}}=\left.\frac{1}{2} r^{3} a
\dot{a}^{2}\right|_{\Sigma}=\frac{1}{2}R(T)\dot{R}^{2}(T)
\end{equation}
The apparent horizon is defined as
\begin{equation}\label{collapse3.9}
R,_{\alpha} R,_{\beta} g ^{\alpha \beta}=0,~~{ i.e.,}~~
r^{2}\dot{a}^{2}=1
\end{equation}
So if $T=T_{AH}$ be the time when the whole cloud starts to be
trapped then
\begin{equation}\label{collapse3.10}
\left. \dot{R}^{2}(T_{AH})\right|_{\Sigma}
=r_{\Sigma}^{2}\dot{a}^{2}(T_{AH})=1
\end{equation}
As it is usually assumed that the collapsing process starts from
regular initial data so initially at $t=t_{i} ~(<T _{AH})$, the
cloud is not trapped i.e.,
\begin{equation}\label{collapse3.11}
r_{\Sigma}^{2} \dot{a}^{2}(t_{i})< 1,~~~~
 (r_{\Sigma}\dot{a}(t_{i})>-1)
\end{equation}
Thus if equation (\ref{collapse3.10}) has any real solution for
$T_{AH}$ satisfying (\ref{collapse3.11}) then black hole(BH) will
form, otherwise the collapsing process leads to a naked
singularity(NS). So the gravitational collapse and consequently
the formation of a BH solely depends upon the nature of root
obtained from equation (\ref{collapse3.10}). If any real solution
for $T_{AH}$ exists for equation (\ref{collapse3.10}) then
apparent horizon will be formed and thus a BH. If there is no real
solution the collapse is destined to result in a NS.

\section{Gravitational Collapse in DGP brane Scenario}\label{chap03}

A simple and effective model of brane-gravity is the
Dvali-Gabadadze-Porrati (DGP) braneworld model \cite{Dvali1,
Deffayet1, Deffayet2} which models our 4-dimensional world as a
FRW brane embedded in a 5-dimensional Minkowski bulk. It explains
the origin of DE as the gravity on the brane leaking to the bulk
at large scale. On the 4-dimensional brane the action of gravity
is proportional to $M_p^2$ whereas in the bulk it is proportional
to the corresponding quantity in 5-dimensions. The model is then
characterized by a cross over length scale $
r_c=\frac{M_p^2}{2M_5^2} $ such that gravity is 4-dimensional
theory at scales $a<<r_c$ where matter behaves as pressureless
dust, but gravity leaks out into the bulk at scales $a>>r_c$ and
matter approaches the behaviour of a cosmological constant.
Moreover it has been shown that the standard Friedmann cosmology
can be firmly embedded in DGP brane.

The Friedmann equation in DGP brane model is modified to the
equation
\begin{equation}\label{collapse3.12}
H^2=\left(\sqrt{\frac{\rho_{T}}{3}+\frac{1}{4r_{c}^{2}}}+\epsilon
\frac{1}{2r_c}\right)^2
\end{equation}
and
\begin{equation}\label{collapse3.13}
\left(2H-\frac{\epsilon}{r_c}\right)\dot{H}=-H\left(\rho_{T}+p_{T}\right),
\end{equation}
where $H=\frac{\dot a}{a}$ is the Hubble parameter, $\rho_{T}$ and
$p_{T}$ are the total cosmic fluid energy density and pressure
respectively and $r_c=\frac{M_p^2}{2M_5^2}$ is the cross-over
scale which determines the transition from 4D to 5D behaviour and
$\epsilon=\pm 1 $ (choosing $M_{p}^{2}=8\pi G=1$). For
$\epsilon=+1$, we have standard DGP$(+)$ model which is self
accelerating model without any form of DE, and effective $w$ is
always non-phantom. However for $\epsilon=-1$, we have DGP$(-)$
model which does not self accelerate but requires DE on the brane.

In this section the role of DM and DE will be shown separately
during the collapse. Then we will study the effect of the
combination of DE and DM in the collapsing process, both in the
presence and in the absence of interaction.

\subsection{Collapse with Dark Matter}
\begin{figure}
\includegraphics[height=2in]{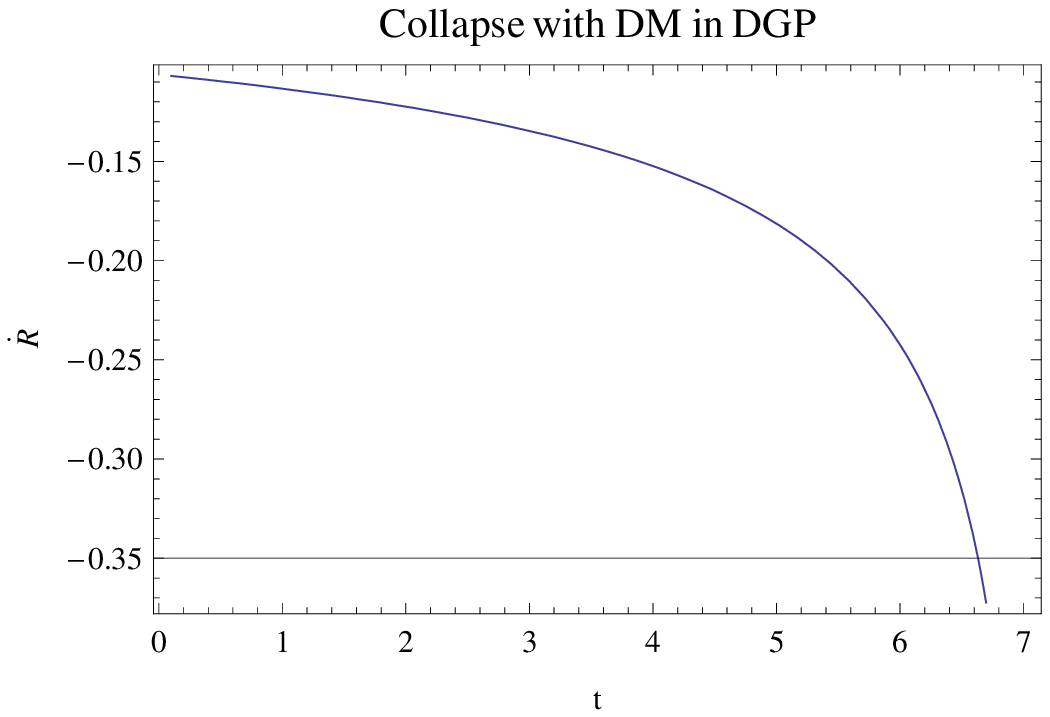}~~~~~~~~\includegraphics[height=2in]{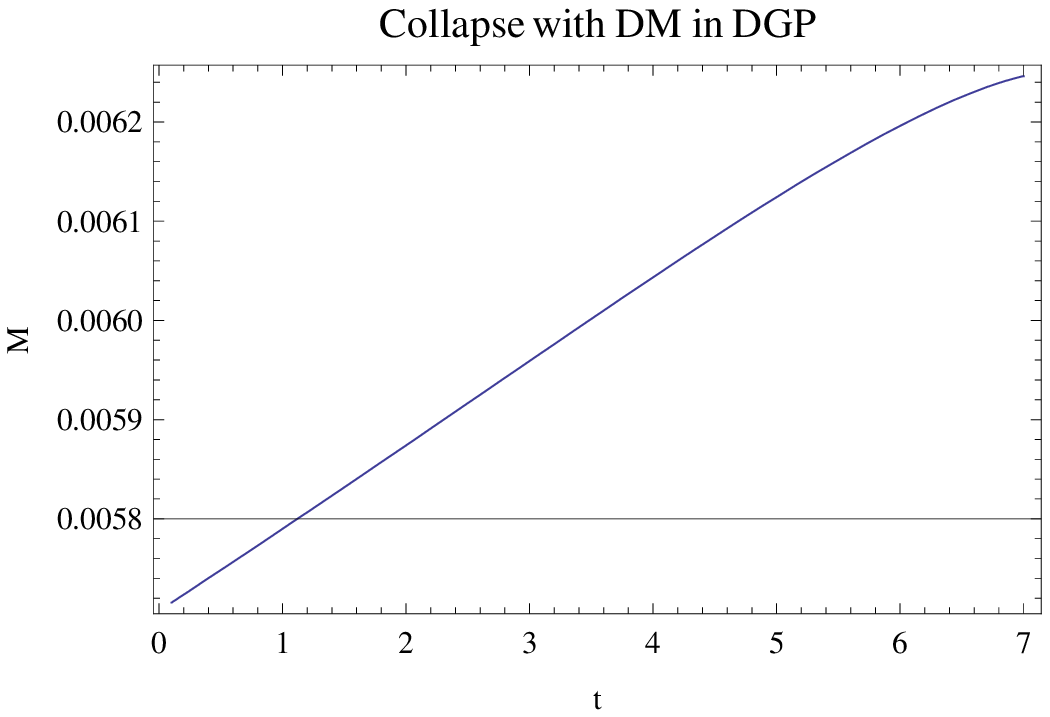}~~\\
\vspace{1mm}
~~~~~~~~~~~Fig. 1~~~~~~~~~~~~~~~~~~~~~~~~~~~~~~~~~~~~~~~~~~~~~~~~~~~~~~~~~~~~~~~~~~Fig. 2~~~~~~~~~~~~~~\\
\vspace{1mm}
\\Fig 1 : The time derivative of the radius is plotted against
time. $\alpha=0.5,~ r=10,~ r_{c}=100$ is considered.
\\Fig 2 : The mass of the collapsing cloud is plotted against
time. $\alpha=0.5,~ r=10,~ r_{c}=100$ is considered.
\end{figure}
Here $\rho _{M}\neq 0, ~\rho_{E} =p_{E} =0$. So from conservation
equation (\ref{collapse3.5}), we have
\begin{equation}\label{collapse3.14}
\rho_{M}=\frac{C_{0}}{a^{3}}
\end{equation}
where $C_{0}$ is the constant of integration. Using this relation
in equation (\ref{collapse3.12}) we get, for DGP(+) model: i.e.,
for $\epsilon=+1$
\begin{equation}\label{collapse3.15}
\frac{2}{3\alpha}\left[\frac{\sqrt{\alpha+a^{3}}}{2}a^{\frac{3}{2}}+\frac{\alpha}{2}\log\left(a^{\frac{3}{2}}+\sqrt{\alpha+a^{3}}\right)-\frac{a^{3}}{2}\right]=-\frac{t}{2r_{c}}+D_{0}
\end{equation}
The corresponding expression for DGP(-) model: i.e., for
$\epsilon=-1$ is,
\begin{equation}\label{collapse3.16}
\frac{2}{3\alpha}\left[\frac{\sqrt{\alpha+a^{3}}}{2}a^{\frac{3}{2}}+\frac{\alpha}{2}\log\left(a^{\frac{3}{2}}+\sqrt{\alpha+a^{3}}\right)+\frac{a^{3}}{2}\right]=-\frac{t}{2r_{c}}+D_{0}
\end{equation}
Where $D_{0}$ is the integration constant
and~~$\alpha=\frac{4r_{c}^{2}C_{0}}{3}$,~~~ Considering DGP(-)
model and using the value of $\rho_{M}$ from
equation(\ref{collapse3.14}) and using the relation
$\rho_{T}=\rho_{M}+\rho_{E}$, the expressions for the time
derivative of geometrical radius $\dot{R}$ and mass $M(T)$ of the
collapsing object are obtained as follows,
\begin{equation}\label{collapse3.17}
\dot{R}(T)=-\frac{r_{\Sigma}\left(\sqrt{\alpha+a^{3}}-a^{3/2}
\right)}{2\sqrt{a}~ r_{c}}
\end{equation}
and
\begin{equation}\label{collapse3.18}
M(T)=\frac{r_{\Sigma}^{3}\left(\sqrt{\alpha+a^{3}}-a^{3/2}
\right)^{2}}{8 r_{c}^{2}}
\end{equation}
From the above solutions it is evident that as~~
$T\rightarrow\infty,~~a\rightarrow\infty,~~\rho_{M}\rightarrow0,~~\dot{R}\rightarrow0,~~M(T)\rightarrow0$.
So we see that there is a tendency of matter density being
diminished as time passes, and finally it tends towards zero. The
time for formation of apparent horizon is given by the real root
of the equation, $\left. \dot{R}^{2}(T_{AH})\right|_{\Sigma}
=r_{\Sigma}^{2}\dot{a}^{2}(T_{AH})=1$ as given by equation
(\ref{collapse3.10}). Thus the corresponding expression for DGP
brane is,
\begin{equation}\label{collapse3.19}
r_{\Sigma}^{2}\left(\sqrt{\alpha+a^{3}}-a^{3/2} \right)^{2}=4a
r_{c}^{2}
\end{equation}
The cloud initially starts untrapped, and gradually as $T$ becomes
equal to $T_{AH}$, the apparent horizon starts forming and the
cloud begins to be shrouded by the horizon. The result is the
formation of a BH singularity.

Normal collapse of matter influences every collapsing particle to
move towards the collapsing centre. So for a normal matter it is
quite natural to have a negative outward velocity at a particular
point. With time as the collapse centre gathers more mass, the
increasing gravitational attraction pulls the rest particles more
efficiently causing an increasing negative outward velocity. The
same here seems to happen for the case of DM. As we do not have
any previous idea of the time gradient of DM collapse. So we
cannot comment whether this is normal or something different.

\subsection{Collapse with Dark Energy in the form of Modified Chaplygin gas}
\begin{figure}
\includegraphics[height=2in]{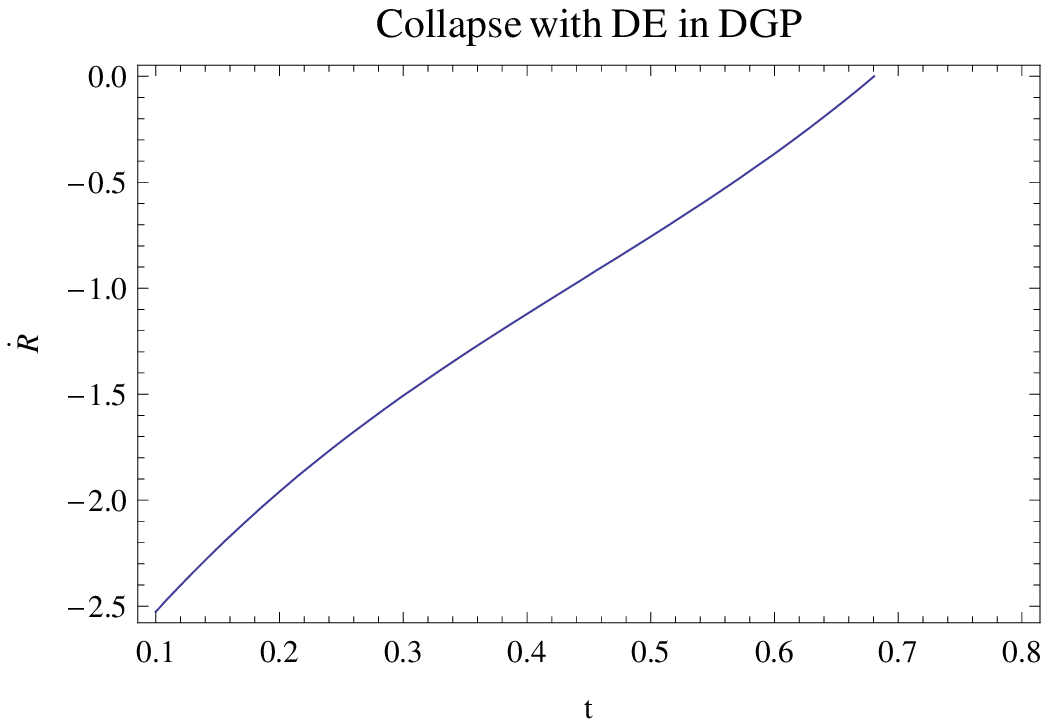}~~~~~~~~\includegraphics[height=2in]{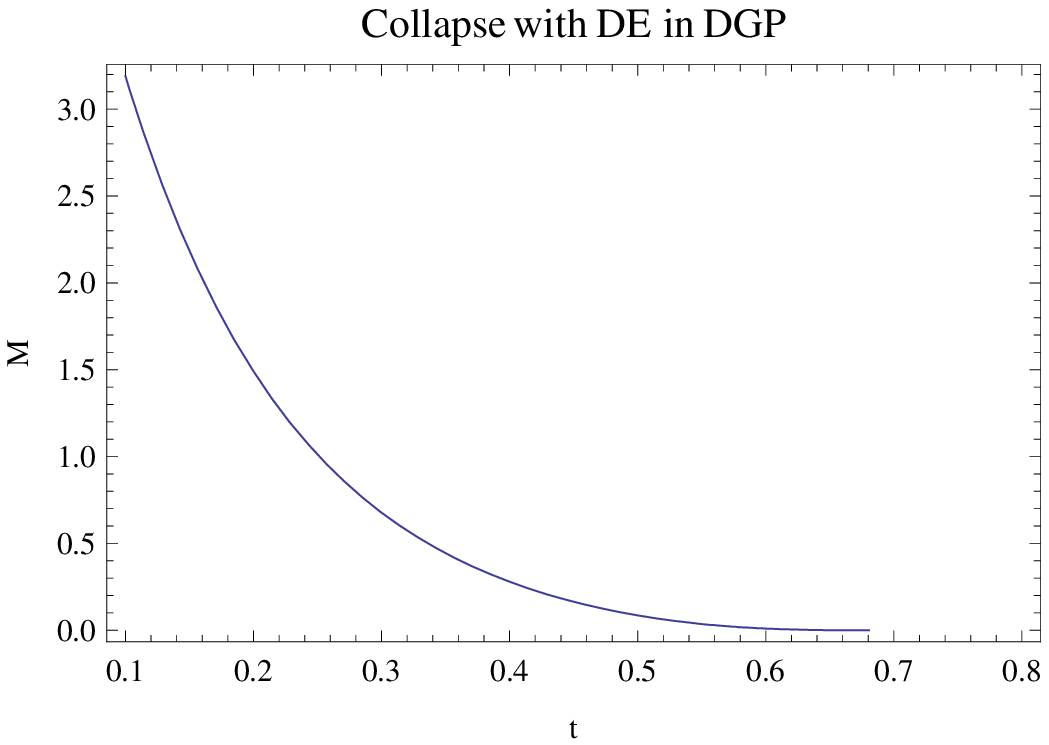}~~\\
\vspace{1mm}
~~~~~~~~~~~Fig. 3~~~~~~~~~~~~~~~~~~~~~~~~~~~~~~~~~~~~~~~~~~~~~~~~~~~~~~~~~~~~~~~~~~Fig. 4~~~~~~~~~~~~~~\\
\vspace{1mm}
\\Fig 3 : The time derivative of the radius is plotted against
time. $r=10,~ r_{c}=1000,~ C_{1}=0.05,~ X_{1}=40,~ X_{2}=0.04,~
X_{3}=0.4 $ is considered.
\\Fig 4 : The mass of the collapsing cloud is plotted against
time. $r=10,~ r_{c}=1000,~ C_{1}=0.05,~ X_{1}=40,~ X_{2}=0.04,~
X_{3}=0.4 $ is considered.
\end{figure}
DE in the form of MCG is considered in this section. Here,
$\rho_{M}=0,~~~p_{MCG}=A\rho_{MCG}-\frac{B}{\rho_{MCG}^{\alpha}}$.
So the solution for density is given by

\begin{equation}\label{collapse3.20}
\rho_{MCG}=\left[\frac{B}{1+A}+\frac{C_{1}}{a^{3(1+A)(1+\alpha)} }
\right]^{\frac{1}{1+\alpha}}
\end{equation}

The expressions for relevant physical quantities are

\begin{equation}\label{collapse3.21}
\dot{R}(T)=-r_{\Sigma}~a~\left[\sqrt{\frac{1}{3}\left(X_{1}+\frac{C_{1}}{a^{X_{2}}}
\right)^{2X_{3}}+\frac{1}{4r_{c}^{2}} } +\frac{\epsilon}{2r_{c}}
\right]
\end{equation}
and
\begin{equation}\label{collapse3.22}
M(T)=\frac{1}{2}r_{\Sigma}^{3}a^{3}
\left[\sqrt{\frac{1}{3}\left(X_{1}+\frac{C_{1}}{a^{X_{2}}}
\right)^{2X_{3}}+\frac{1}{4r_{c}^{2}} } +\frac{\epsilon}{2r_{c}}
\right]^{2}
\end{equation}

~~~~~~~~where~~~~$X_{1}=\frac{B}{1+A}$,~~~~~$X_{2}=3\left(1+A\right)\left(1+\alpha\right)$,~~~~~$X_{3}=\frac{1}{2\left(1+\alpha\right)}.~~~~~$
The limiting values of the physical parameters are as follows:
$$\textbf{case1}$$
$$~~~~~a\rightarrow0:~~~~\rho_{MCG}\rightarrow\infty,~~~~for ~~~1+A>0;~~~~\rho_{MCG}\rightarrow\left[\frac{B}{1+A}\right]^{\frac{1}{1+\alpha}},~~~~for~~~~1+A<0$$
$$~~~~\dot{R}\rightarrow\frac{-rC_{1}^{\frac{1}{2(1+\alpha)}}a^{\frac{-(1+3A)}{2}}}{\sqrt{3r_{c}}},~~~for~~~1+A>0;~~~~\dot{R}\rightarrow-\infty,~~~~~~~ for~~~~~~~~1+A<0$$
$$~~~~M(T)\rightarrow\frac{r^{3}a^{-3A}C_{1}^{\frac{1}{1+\alpha}}}{6},~~~~for~~~~1+A>0;~~~~M(T)\rightarrow\infty,~~~~~for~~~~1+A<0$$

$$\textbf{case2}$$
$$~~~~a\rightarrow\infty:~~~~\rho_{MCG}\rightarrow\left[\frac{B}{1+A}\right]^{\frac{1}{1+\alpha}},~~~~for ~~~1+A>0;~~~~\rho_{MCG}\rightarrow\infty,~~~~for ~~~1+A<0 $$
$$~~~~\dot{R}\rightarrow-\infty,~~~~for~~~~1+A>0;~~~~\dot{R}\rightarrow\frac{-rC_{1}^{\frac{1}{2(1+\alpha)}}a^{\frac{-(1+3A)}{2}}}{\sqrt{3r_{c}}},~~~~~for~~~~1+A<0$$
$$~~~~M(T)\rightarrow\infty,~~~~~for~~~~1+A>0;~~~~M(T)\rightarrow\frac{r^{3}a^{-3A}C_{1}^{\frac{1}{1+\alpha}}}{6},~~~~for~~~1+A<0$$
The limiting value of $M(T)$ for $1+A>0$ is unphysical for
$a\rightarrow\infty$. This is because the mass of the collapsing
cloud has to decrease in presence of MCG due to accretion. The
cloud will start untrapped at the instant given by the real roots
of the following equation and gradually start to be trapped,
\begin{equation}\label{collapse3.23}
r_{\Sigma}^{2}~a^{2}~\left[\sqrt{\frac{1}{3}\left(X_{1}+\frac{C_{1}}{a^{X_{2}}}
\right)^{2X_{3}}+\frac{1}{4r_{c}^{2}} } +\frac{\epsilon}{2r_{c}}
\right]^{2}=1
\end{equation}

In Fig.3 and Fig.4 we find the time gradient for R, i.e.,
$\dot{R}$ and mass against time for DE collapse. Here we can see
that at first the DE collapse requires an outward negative
velocity. However, $\dot{R}$ has a monotonically decreasing
magnitude which tends to zero with increasing time. This implies,
with time the DE collapse decreases, the tendency to collapse.
Even the final fate might be the fact that there is no ingoing
fluid. The change of mass is quite amazing in nature as it sharply
indicates that matter is flowing outward from collapsing centre.

\subsection{Collapse with Dark Energy in the form of Generalised Cosmic Chaplygin Gas}
\begin{figure}
\includegraphics[height=2in]{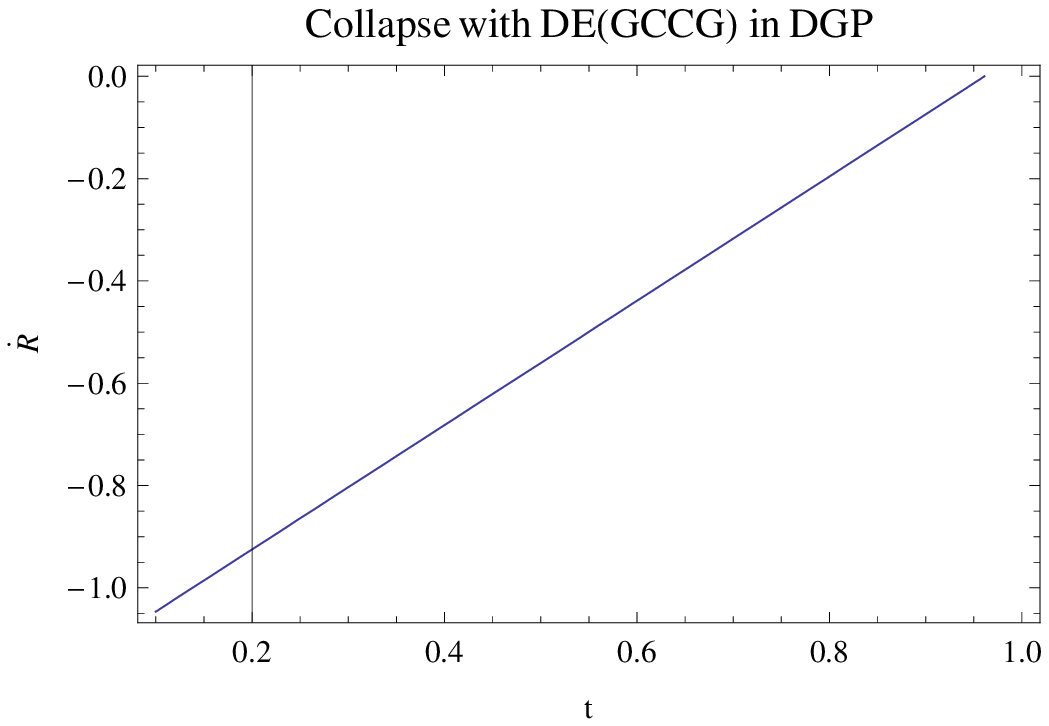}~~~~~~~~\includegraphics[height=2in]{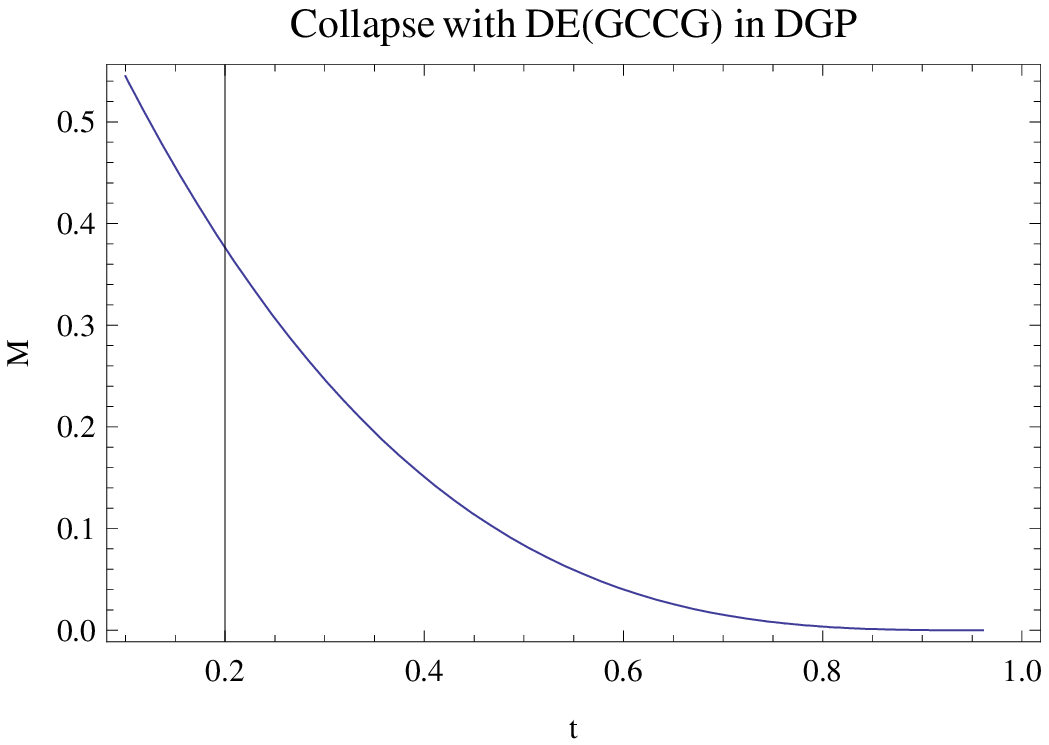}~~\\
\vspace{1mm}
~~~~~~~~~~~Fig. 5~~~~~~~~~~~~~~~~~~~~~~~~~~~~~~~~~~~~~~~~~~~~~~~~~~~~~~~~~~~~~~~~~~Fig. 6~~~~~~~~~~~~~~\\
\vspace{1mm}
\\Fig 5 : The time derivative of the radius is plotted against
time. $r=10,~ r_{c}=100000,~ C'=10, ~B'=5, ~ X_{1}=0.00005,~
X_{2}=1.5,~ X_{3}=0.25 $ is considered.
\\Fig 6 : The mass of the collapsing cloud is plotted against
time. $r=10,~ r_{c}=100000,~ C'=10, ~B'=5, ~ X_{1}=0.00005,~
X_{2}=1.5,~ X_{3}=0.25 $ is considered.
\end{figure}
Here DE in the form of GCCG is considered. So,
$\rho_{M}=0$,~~$p=-\rho_{GCCG}^{-\alpha}\left[C'+\left(\rho_{GCCG}^{1+\alpha}-C'\right)^{-\omega}\right]$
as given in equation (\ref{collapse3.2}). The solution for density
is given by
\begin{equation}\label{collapse3.24}
\rho_{GCCG}=\left[C'+\left\{1+\frac{B'}{a^{3(1+\alpha)(1+\omega)}}\right\}^\frac{1}{1+\omega}\right]^\frac{1}{1+\alpha}~~,~~B'~
is~ positive~ integration~ constant.
\end{equation}
The expressions for relevant physical quantities are
\begin{equation}\label{collapse3.25}
\dot{R}(T)=-r_{\Sigma}~a~\left[\sqrt{\frac{1}{3}\left[C'+\left\{1+\frac{B'}{a^{X_{2}'}}\right\}^{X_{1}'}\right]^{2X_{3}'}~+\frac{1}{4r_{c}^{2}}
} +\frac{\epsilon}{2r_{c}} \right]
\end{equation}
and
\begin{equation}\label{collapse3.26}
M(T)=\frac{1}{2}r_{\Sigma}^{3}a^{3}
\left[\sqrt{\frac{1}{3}\left[C'+\left\{1+\frac{B'}{a^{X_{2}'}}\right\}^{X_{1}'}\right]^{2X_{3}'}~+\frac{1}{4r_{c}^{2}}
} +\frac{\epsilon}{2r_{c}} \right]^{2}
\end{equation}
where
$X_{1}'=\frac{1}{1+\omega}$,~~~$X_{2}'=3\left(1+\alpha\right)\left(1+\omega\right)$,~~~$X_{3}'=\frac{1}{2\left(1+\alpha\right)}.$
~~ The limiting values of the physical parameters are as follows:
~~~~~~~~~~~~~~~~~~~~~~~~~~~~~~~~~~~~~~~~~~~~~~~~~~~~~~~~~~~~~~~~~~~~~~~~~~~~~~~~~~~~~~~~~~~~~~~~~~~~~~~~~~~~~~~~~~~~~~~~~~~~~~~~~~~~~~~~~~~~~~~~~~~~~~~~~~~~~~~~~~~~~~~~~
~~~~~~~~~~~~~~~~~~~~~~~~~~~~~~~~~~~~~~~~~~~~~~~~~~~~~~~~~~~~~~~~~~~~~~~~~~~~~~~~~~~~~~~~~~~~~~~~~~~~~~~~~~~~~~~~~~~~~~~~~~~~~~~~~~~~~~~~~~~~~~~~~~~~~~~~~~~~~~~~~~~~~~~~~~~~~~~~~~~~~~~~~~~
$$\textbf{case1}$$
$$~~~~~a\rightarrow0:~~~~\rho_{GCCG}\rightarrow\infty,~~~~for ~~~1+\omega>0;~~~~\rho_{GCCG}\rightarrow\left(C'+1\right)^{\frac{1}{1+\alpha}},~~~~for~~~~1+\omega<0$$
$$~~~~\dot{R}\rightarrow-\frac{r}{\sqrt{3a}}B'^{\frac{1}{2(1+\alpha)(1+\omega)}},~~~for~~~1+\omega>0$$
$$~~~~M(T)\rightarrow\frac{r^{3}}{6a^{2}}B'^{\frac{1}{(1+\alpha)(1+\omega)}},~~~~for~~~~1+\omega>0$$

$$\textbf{case2}$$
$$~~~~~a\rightarrow\infty:~~~~\rho_{GCCG}\rightarrow\infty,~~~~for ~~~1+\omega<0;~~~~\rho_{GCCG}\rightarrow\left(C'+1\right)^{\frac{1}{1+\alpha}},~~~~for~~~~1+\omega>0$$
$$~~~~\dot{R}\rightarrow-\frac{ra}{2\sqrt{3}r_{c}}\left(\sqrt{4r_{c}^{2}C'^{\frac{1}{1+\alpha}}+3}+\sqrt{3}\epsilon\right),~~~~for~~~~1+\omega>0$$
$$~~~~M(T)\rightarrow\frac{ar^{3}}{24r_{c}^{2}}\left(\sqrt{4r_{c}^{2}\left(C'+1\right)^{\frac{1}{1+\alpha}}+3}+\sqrt{3}\epsilon\right)^{2},~~~~~for~~~~1+\omega>0$$
The corresponding expression for the formation of event horizon
is,
\begin{equation}\label{collapse3.27}
r_{\Sigma}^{2}~a^{2}~\left[\sqrt{\frac{1}{3}\left[C'+\left\{1+\frac{B'}{a^{X_{2}'}}\right\}^{X_{1}'}\right]^{2X_{3}'}~+\frac{1}{4r_{c}^{2}}
} +\frac{\epsilon}{2r_{c}} \right]^{2}=1
\end{equation}

Fig.5 and Fig.6 are the graphs for $\dot{R}$ and $M$ respectively
for GCCG collapse. They almost resemble with Fig.3 and Fig.4
respectively. Only the $\dot{R}$ curve is almost a straight line
whereas in case of MCG, it was a curved one.

\subsection{Effect of a combination of dark matter and dark energy(in the form of MCG)}
\subsubsection{Case I : $Q = 0$ i.e., No Interaction Between Dark Matter And Dark Energy:}
\begin{figure}
\includegraphics[height=2in]{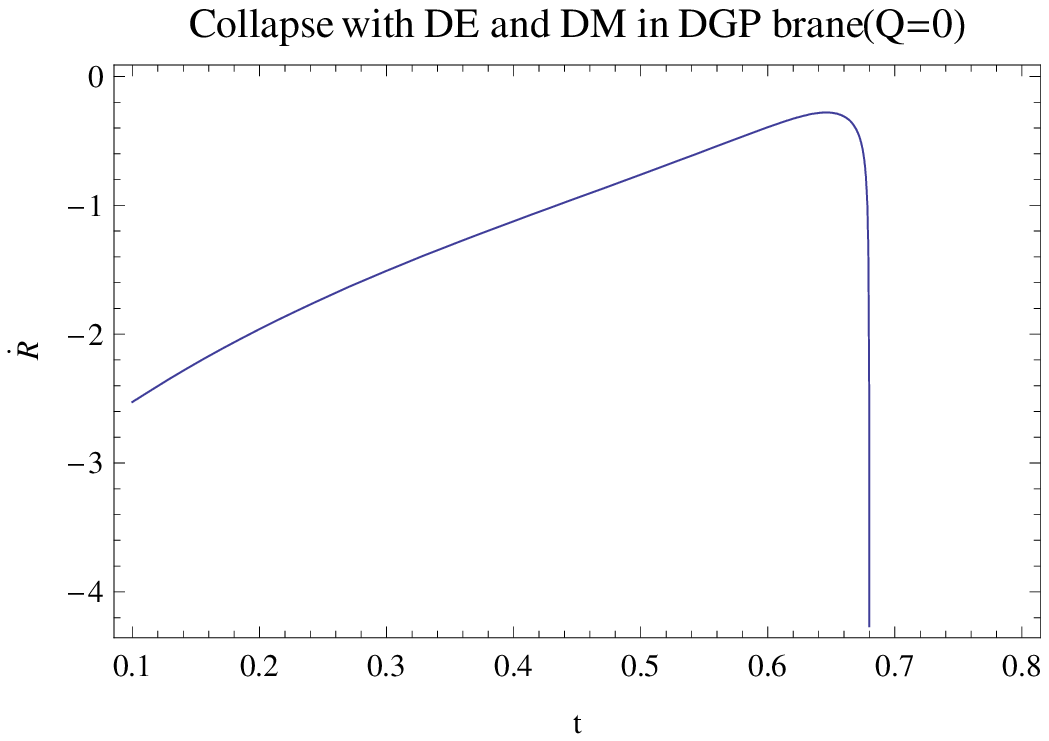}~~~~~~~~\includegraphics[height=2in]{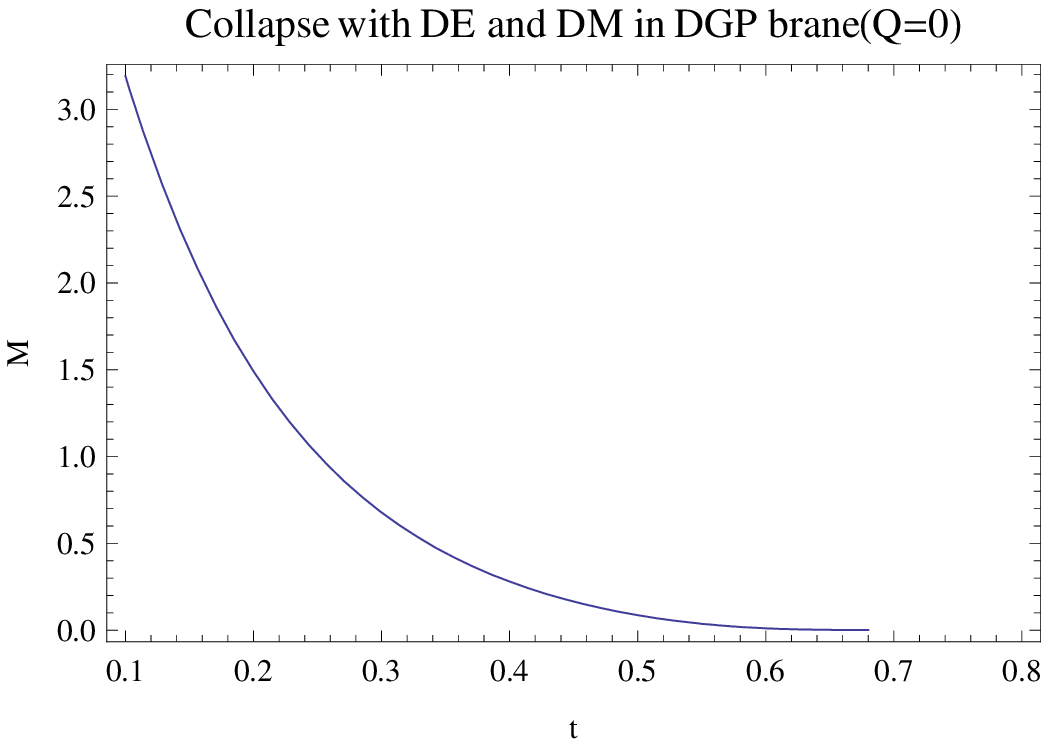}~~\\
\vspace{1mm}
~~~~~~~~~~~Fig. 7~~~~~~~~~~~~~~~~~~~~~~~~~~~~~~~~~~~~~~~~~~~~~~~~~~~~~~~~~~~~~~~~~~Fig. 8~~~~~~~~~~~~~~\\
\vspace{1mm}
\\Fig 7 : The time derivative of the radius is plotted against
time. $r=10,~ r_{c}=1000,~ C_{0}=0.00001, ~C_{1}=0.05, ~
X_{1}=40,~ X_{2}=0.04,~ X_{3}=0.4 $ is considered.
\\Fig 8 : The mass of the collapsing cloud is plotted against
time. $r=10,~ r_{c}=1000,~ C_{0}=0.00001, ~C_{1}=0.05, ~
X_{1}=40,~ X_{2}=0.04,~ X_{3}=0.4 $ is considered.
\end{figure}
\begin{figure}
\includegraphics[height=2in]{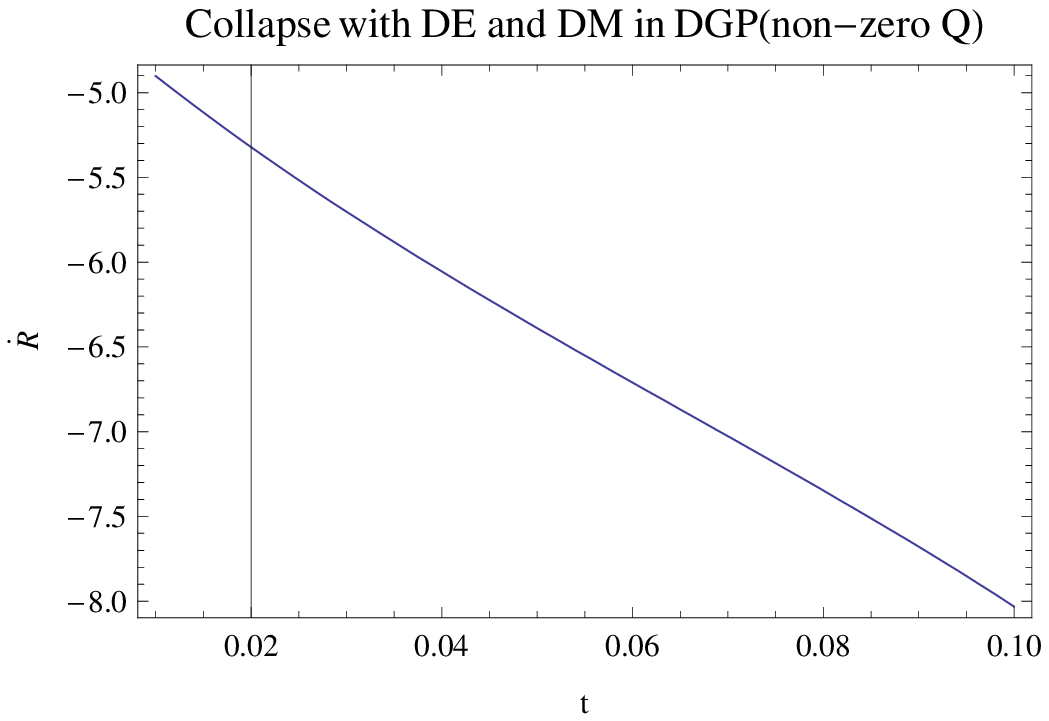}~~~~~~~~\includegraphics[height=2in]{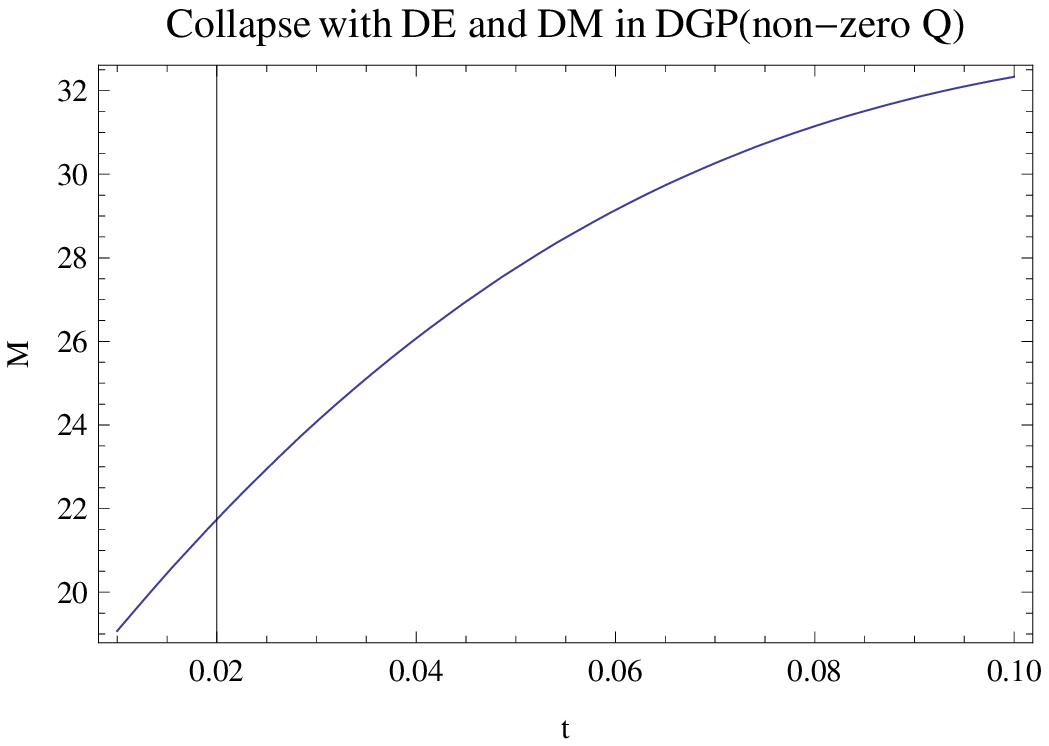}~~\\
\vspace{1mm}
~~~~~~~~~~~Fig. 9~~~~~~~~~~~~~~~~~~~~~~~~~~~~~~~~~~~~~~~~~~~~~~~~~~~~~~~~~~~~~~~~~~Fig. 10~~~~~~~~~~~~~~\\
\vspace{1mm}
\\Fig 9 : The time derivative of the radius is plotted against
time. $r=10,~ r_{c}=1000,~n=1,~A=1,~B=1000,~\alpha=0.5,~z_{0}=1 $
is considered.
\\Fig 10 : The mass of the collapsing cloud is plotted against
time. $r=10,~ r_{c}=1000,~n=1,~A=1,~B=1000,~\alpha=0.5,~z_{0}=1$
is considered.
\end{figure}
Energy density of matter as given by equation (\ref{collapse3.14})
is~ $\rho_{M}=\frac{C_{0}}{a^{3}}$ . Also the energy density of
MCG is given in equation (\ref{collapse3.20}). The expressions for
relevant physical quantities are
\begin{equation}\label{collapse3.28}
\dot{R}(T)=-r_{\Sigma}~a~\left[\sqrt{\frac{1}{3}\left\{\frac{C_{0}}{a^{3}}+
\left(X_{1}+\frac{C_{1}}{a^{X_{2}}}
\right)^{2X_{3}}\right\}+\frac{1}{4r_{c}^{2}} }
+\frac{\epsilon}{2r_{c}} \right]
\end{equation}
and
\begin{equation}\label{collapse3.29}
M(T)=\frac{1}{2}r_{\Sigma}^{3}a^{3}
\left[\sqrt{\frac{1}{3}\left\{\frac{C_{0}}{a^{3}}+\left(X_{1}+\frac{C_{1}}{a^{X_{2}}}
\right)^{2X_{3}}\right\}+\frac{1}{4r_{c}^{2}} }
+\frac{\epsilon}{2r_{c}} \right]^{2}
\end{equation}

~~~~~The limiting values of the physical parameters are as
follows,
$$\textbf{case1}$$
$$~~~~~a\rightarrow0:~~~\rho_{M}\rightarrow\infty,~~~~~\rho_{MCG}\rightarrow\infty,~~~~for ~~~1+A>0;~~~~~\rho_{MCG}\rightarrow\left[\frac{B}{1+A}\right]^{\frac{1}{1+\alpha}},~~~~~for ~~~~1+A<0 $$
$$~~~~~\dot{R}\rightarrow\frac{-\epsilon ar}{2r_{c}},~~~~for ~~~1+A>0;~~~~M(T)\rightarrow\frac{C_{1}^{\frac{1}{1+\alpha}}a^{-3A}r^{3}}{6},~~~~for~~~~1+A>0$$

 $$\textbf{case2}$$
$$~~~~~a\rightarrow\infty:~~~~~\rho_{M}\rightarrow0,~~~~\rho_{MCG}\rightarrow\left[\frac{B}{1+A}\right]^{\frac{1}{1+\alpha}},~~~~for ~~~1+A>0;~~~~~\rho_{MCG}\rightarrow\infty,~~~~for ~~~1+A<0 $$
$$~~~~~\dot{R}\rightarrow\frac{-\sqrt{2}ra^{-(1+3A)}C_{1}^{\frac{1}{2(1+\alpha)}}}{3},~~~~~ for~~~~1+A<0;~~~~M(T)\rightarrow\frac{C_{1}^{\frac{1}{1+\alpha}}a^{-3A}r^{3}}{6},~~~~~for~~~~1+A<0$$

\subsubsection{Case II : $Q \neq 0$ i.e., Interaction Between
Dark Matter And Dark Energy:}
In this case walking in the path of Cai and Wang \cite{Cai1, Cai2}
we will assume the relation
\begin{equation}\label{collapse3.30}
\frac{\rho_{MCG}}{\rho_{M}}=Ca^{3n}
\end{equation}
where $C>0$ and n are arbitrary constants. Now we solve the
conservation equations (\ref{collapse3.5}) and (\ref{collapse3.6})
to get the following expression for $\rho_{T}$ where
$\rho_{T}=\rho_{MCG}+\rho_{M}$,

$$\rho_{T}^{\alpha+1}=\frac{\left(\alpha+1\right)B}{\left[\alpha\left(n-1\right)-1\right]}\frac{\left(Ca^{3n}\right)^{\frac{2}{n}\left(\alpha+1-n\alpha\right)-\left(\alpha+1\right)}}{\left(Ca^{3n}+1\right)^{\frac{2}{n}\left(\alpha+1-n\alpha\right)+A\left(\alpha+1\right)}}$$
$$~~~~~~_{2}F_{1}\left[\frac{1+\alpha-n\alpha}{n},\frac{1+n+\alpha+A+A\alpha}{n},\frac{1+n+\alpha-n\alpha}{n},\frac{Ca^{3n}}{1+Ca^{3n}}\right]$$
\begin{equation}\label{collapse3.31}~~~~~~+z_{0}\left[Ca^{3n}\left(1+Ca^{3n}\right)^{A}\right]^{-\left(\alpha+1\right)}
\end{equation}
where $_{2}F_{1}$ is the hypergeometric function and $z_{0}$ is
the integration constant. Using the relation
$\rho_{T}=\rho_{MCG}+\rho_{M}$ we get
\begin{equation}\label{collapse3.32}
\rho_{MCG}=\frac{Ca^{3n}\rho_{T}}{1+Ca^{3n}}~~~~~~
~~~~~~~~~~~~~~~~~~\rho_{M}=\frac{\rho_{T}}{1+Ca^{3n}}
\end{equation}

The expression for interaction in this case, is obtained by using
equations \ref{collapse3.1}, \ref{collapse3.5} and
\ref{collapse3.12} and is given by,
\begin{equation}\label{collapse3.33}
Q=-\frac{3\rho_{T}}{1+Ca^{3n}}\left[\sqrt{\frac{\rho_{T}}{3}+\frac{1}{4r_{c}^{2}}}+\frac{\epsilon}{2r_{c}}\right]\left[\frac{1}{1+Ca^{3n}}\left(1+a^{3n}\left(C\left(1+A+3n\right)\right)-BC^{-\alpha}a^{-3\alpha
n}\left(1+Aa^{3n}\right)^{\alpha+1}\rho_{T}^{-\alpha-1}\right)-1\right]
\end{equation}
In this case the value for the gradient of scale factor is given
by

$$\dot{a}=-a\left[\left(\frac{A_{3}}{a^{3n}(1+Ca^{3n})^{A}}\left[\frac{C^{\frac{2}{n}(\alpha+1-n\alpha)}a^{6(\alpha+1-n\alpha)}}{(1+Ca^{3n})^{\frac{2}{n}(\alpha+1-n\alpha)}}~_{2}F_{1}\left[\frac{1+\alpha-n\alpha}{n},\frac{1+n+\alpha+A+A\alpha}{n},\frac{1+n+\alpha-n\alpha}{n},\right.\right.\right.\right.$$
\begin{equation}\label{collapse3.34}
\left.\left.\left.\left.\frac{Ca^{3n}}{1+Ca^{3n}}\right]+z_{1}\right]^{\frac{1}{\alpha+1}}+\frac{1}{4r_{c}^{2}}\right)^{\frac{1}{2}}+\frac{\epsilon}{2r_{c}}\right]
\end{equation}
where
$A_{3}=\frac{1}{3C}\left[\frac{(\alpha+1)B}{\alpha(n-1)-1}\right]^{\frac{1}{1+\alpha}}$,
~~~~~and~~~~~$z_{1}=z_{0}\left[\frac{(\alpha+1)B}{\alpha(n-1)-1}\right]^{\frac{-1}{\alpha+1}}$.~~
The corresponding expressions for $\dot{R}$ and mass $M(T)$ is
given as follows:

$$\dot{R}=-r_{\Sigma}a\left[\left(\frac{A_{3}}{a^{3n}(1+Ca^{3n})^{A}}\left[\frac{C^{\frac{2}{n}(\alpha+1-n\alpha)}a^{6(\alpha+1-n\alpha)}}{(1+Ca^{3n})^{\frac{2}{n}(\alpha+1-n\alpha)}}~_{2}F_{1}\left[\frac{1+\alpha-n\alpha}{n},\frac{1+n+\alpha+A+A\alpha}{n},\frac{1+n+\alpha-n\alpha}{n},\right.\right.\right.\right.$$
\begin{equation}\label{collapse3.35}
\left.\left.\left.\left.\frac{Ca^{3n}}{1+Ca^{3n}}\right]+z_{1}\right]^{\frac{1}{\alpha+1}}+\frac{1}{4r_{c}^{2}}\right)^{\frac{1}{2}}+\frac{\epsilon}{2r_{c}}\right]
\end{equation}
and

$$M(T)=\frac{1}{2}a^{3}r_{\Sigma}^{3}\left[\left(\frac{A_{3}}{a^{3n}(1+Ca^{3n})^{A}}\left[\frac{C^{\frac{2}{n}(\alpha+1-n\alpha)}a^{6(\alpha+1-n\alpha)}}{(1+Ca^{3n})^{\frac{2}{n}(\alpha+1-n\alpha)}}\right.\right.\right.$$
$$\left.\left.\left._{2}F_{1}\left[\frac{1+\alpha-n\alpha}{n},\frac{1+n+\alpha+A+A\alpha}{n},\frac{1+n+\alpha-n\alpha}{n},\right.\right.\right.\right.$$
\begin{equation}\label{collapse3.36}
\left.\left.\left.\left.\frac{Ca^{3n}}{1+Ca^{3n}}\right]+z_{1}\right]^{\frac{1}{\alpha+1}}+\frac{1}{4r_{c}^{2}}\right)^{\frac{1}{2}}+\frac{\epsilon}{2r_{c}}\right]^{2}
\end{equation}
From the above expressions the limiting behaviour of the physical
parameters are obtained as follows:

$$\textbf{case1}$$
$$ a\rightarrow0:~~~~\rho_{M}\rightarrow~a^{-3n},~~~~\rho_{MCG}\rightarrow~a~ constant,~~~~\rho_{T}\rightarrow~a^{-3n},~~~~\dot{R}\rightarrow~-a^{1-\frac{3n}{2}},~~~~M(T)\rightarrow~a^{3(1-n)}$$

$$\textbf{case2}$$
$$ a\rightarrow~\infty:~~~~\rho_{M}\rightarrow~a^{-3n(\alpha+1)(A+1)},~~~~\rho_{MCG}\rightarrow~a^{-3n(\alpha+1)(A+1)},~~~~\rho_{T}\rightarrow~a^{-3n(\alpha+1)(A+1)},~~~~\dot{R}\rightarrow~-a^{1-\frac{3n}{2}(1+A)}$$
$$M(T)\rightarrow~a^{3[1-n(1+A)]}$$

Fig. 7,8,9 and 10 are for the cases where DM and DE are present
together. First two characterizes the no interaction scenario,
whereas the last two deals with the interaction cases.

When there is no interaction the magnitude of $\dot{R}$ decreases
at first followed by a steep increment. This can be interpreted as
the domination of DE over DM at the first phase and domination of
DM over DE at the latter stage. In either case the mass decreases
for no interaction case and lastly becomes asymptotic with time
axis.

For interacting DM and DE model, it is clear that DM dominates and
the basic nature of the graphs are similar to DM collapse case.

\subsection{Effect of combination of dark matter and dark energy(in the form of GCCG)}
\subsubsection{Case I : $Q = 0$ i.e., No Interaction Between Dark Matter And Dark Energy:}
\begin{figure}
\includegraphics[height=2in]{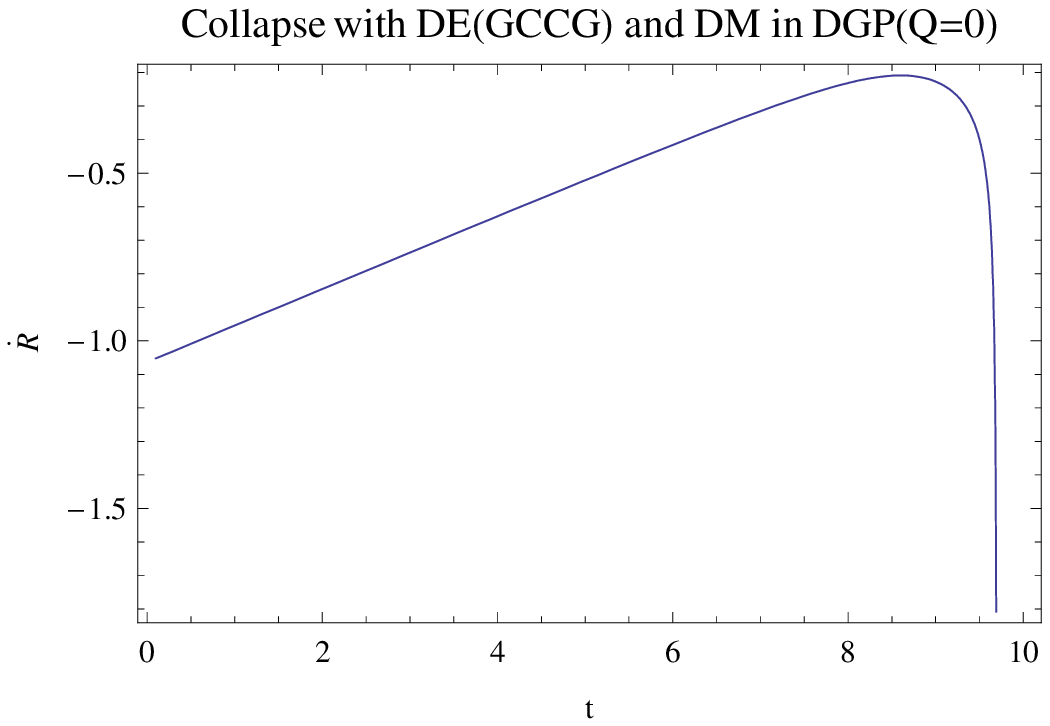}~~~~~~~~\includegraphics[height=2in]{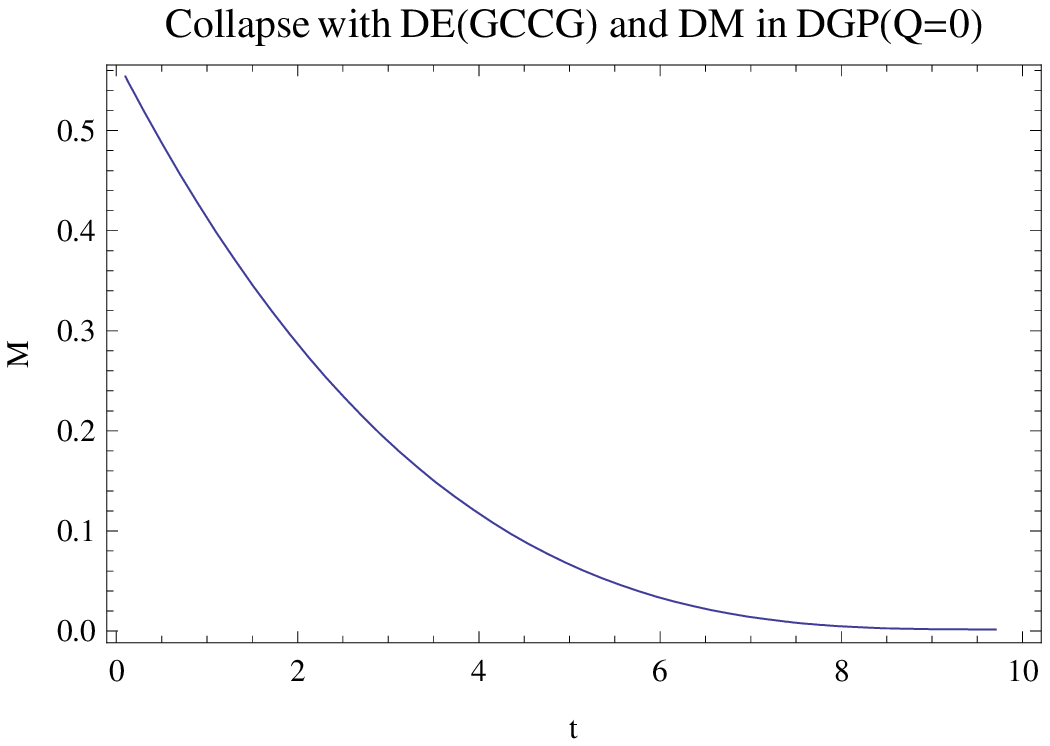}~~\\
\vspace{1mm}
~~~~~~~~~~~Fig. 11~~~~~~~~~~~~~~~~~~~~~~~~~~~~~~~~~~~~~~~~~~~~~~~~~~~~~~~~~~~~~~~~~~Fig. 12~~~~~~~~~~~~~~\\
\vspace{1mm}
\\Fig 11 : The time derivative of the radius is plotted against
time. $r=10,~
r_{c}=1000,~n=1,~C'=10,~C_{0}=0.00001,~B'=5,~X_{1}=0.00005,~X_{2}=1.5,~X_{3}=0.25
$ is considered.
\\Fig 12 : The mass of the collapsing cloud is plotted against
time. $r=10,~
r_{c}=1000,~n=1,~C'=10,~C_{0}=0.00001,~B'=5,~X_{1}=0.00005,~X_{2}=1.5,~X_{3}=0.25$
is considered.
\end{figure}

\begin{figure}
\includegraphics[height=2in]{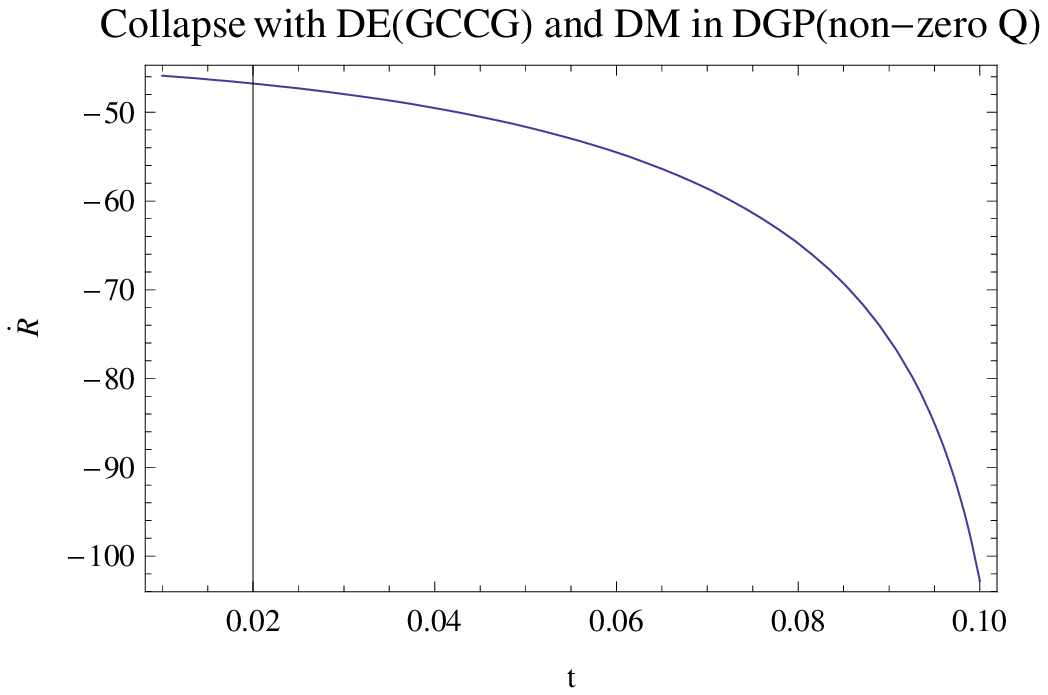}~~~~~~~~\includegraphics[height=2in]{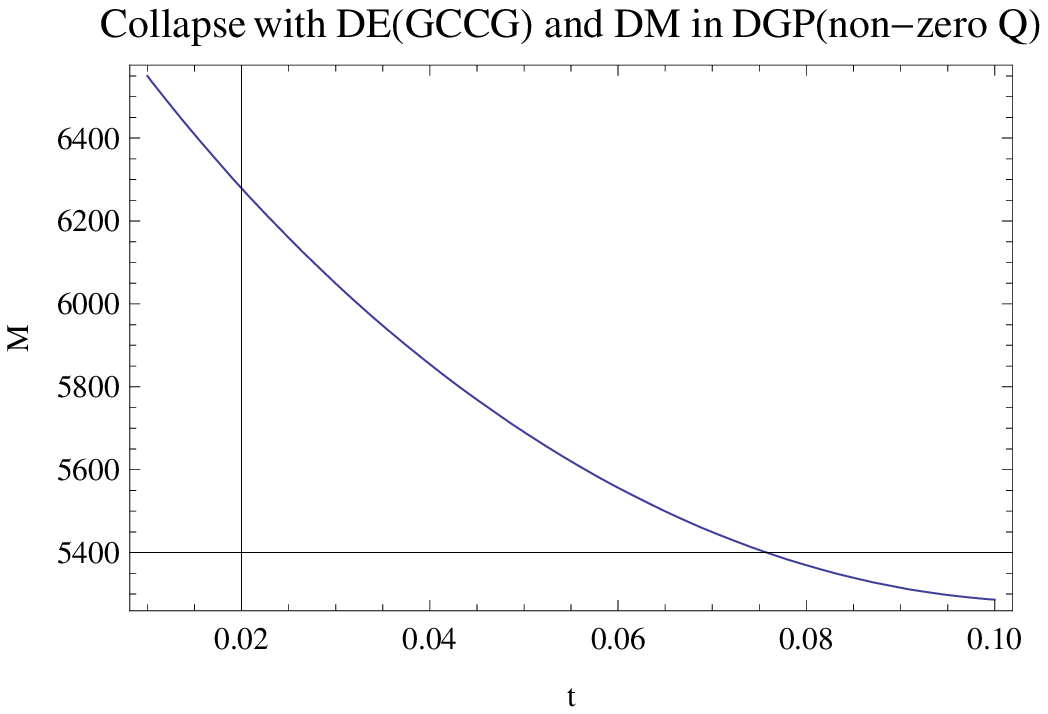}~~\\
\vspace{1mm}
~~~~~~~~~~~Fig. 13~~~~~~~~~~~~~~~~~~~~~~~~~~~~~~~~~~~~~~~~~~~~~~~~~~~~~~~~~~~~~~~~~~Fig. 14~~~~~~~~~~~~~~\\
\vspace{1mm}
\\Fig 13 : The time derivative of the radius is plotted against
time. $r=10,~
r_{c}=100000,~n=1,~\widetilde{C}=0.001,~B'=5,~\alpha=0.5,~C'=1000
$ is considered.
\\Fig 14 : The mass of the collapsing cloud is plotted against
time. $r=10,~
r_{c}=100000,~n=1,~\widetilde{C}=0.001,~B'=5,~\alpha=0.5,~C'=1000$
is considered.
\end{figure}

The energy density of DM is given by equation (\ref{collapse3.14})
and the energy density of GCCG is given by equation
(\ref{collapse3.24}). Using these two equations in the modified
Friedmann equation for DGP brane, i.e.,
equation(\ref{collapse3.12}), we obtain the expressions for the
relevant parameters as given below.

The time derivative of the geometrical radius of the collapsing
star is given by,
\begin{equation}\label{collapse3.37}
\dot{R}(T)=-ar_{\Sigma}\left[\sqrt{\frac{1}{3}\left(\frac{C_{0}}{a^{3}}+\left(C'+\left(1+\frac{B'}{a^{X_{2}'}}\right)^{X_{1}'}\right)^{2X_{3}'}\right)+\frac{1}{4r_{c}^{2}}}+\frac{\epsilon}{2r_{c}}\right]
\end{equation}
The expression for mass of the collapsing cloud is given by,
\begin{equation}\label{collapse3.38}
M(T)=\frac{1}{2}a^{3}r_{\Sigma}^{3}\left[\sqrt{\frac{1}{3}\left(\frac{C_{0}}{a^{3}}+\left(C'+\left(1+\frac{B'}{a^{X_{2}'}}\right)^{X_{1}'}\right)^{2X_{3}'}\right)+\frac{1}{4r_{c}^{2}}}+\frac{\epsilon}{2r_{c}}\right]^{2}
\end{equation}

The limiting values of the physical parameters are given below,
$$\textbf{case1}$$
$$~~~~~a\rightarrow0:~~~~\rho_{GCCG}\rightarrow\infty,~~~~for ~~~1+\omega>0;~~~~\rho_{GCCG}\rightarrow\left(C'+1\right)^{\frac{1}{1+\alpha}},~~~~for~~~~1+\omega<0$$
$$~~~~\dot{R}\rightarrow-\frac{r}{\sqrt{3a}}\sqrt{C_{0}+B'^{\frac{1}{(1+\alpha)(1+\omega)}}},~~~for~~~1+\omega>0$$
$$~~~~M(T)\rightarrow\frac{r^{3}}{6}\left(C_{0}+B'^{\frac{1}{(1+\alpha)(1+\omega)}}\right),~~~~for~~~~1+\omega>0$$

$$\textbf{case2}$$
$$~~~~~a\rightarrow\infty:~~~~\rho_{GCCG}\rightarrow\infty,~~~~for ~~~1+\omega<0;~~~~\rho_{GCCG}\rightarrow\left(C'+1\right)^{\frac{1}{1+\alpha}},~~~~for~~~~1+\omega>0$$
$$~~~~\dot{R}\rightarrow-\frac{r\sqrt{a}}{\sqrt{3}}\left[\sqrt{\left(C'+1\right)^{\frac{1}{1+\alpha}}+\frac{3}{4r_{c}^{2}}}+\frac{\sqrt{3}\epsilon}{2r_{c}}\right],~~~~for~~~~1+\omega>0$$
$$~~~~M(T)\rightarrow\frac{a^{3}r^{3}}{6}\left[\sqrt{\left(C'+1\right)^{\frac{1}{1+\alpha}}+\frac{3}{4r_{c}^{2}}}+\frac{\sqrt{3}\epsilon}{2r_{c}}\right]^{2},~~~~for~~~~1+\omega>0$$

\subsubsection{Case II : $Q \neq 0$ i.e., Interaction Between
Dark Matter And Dark Energy:}
Like the case of MCG here also we assume the relation
\begin{equation}\label{collapse3.39}
\frac{\rho_{GCCG}}{\rho_{M}}=ka^{3n}
\end{equation}
where $k>0$ and n are arbitrary constants. Using this equation and
equation \ref{collapse3.24} in the equations \ref{collapse3.5} and
\ref{collapse3.6}, we calculate the expression for total energy
density ($\rho_{T}$) as given below,

$$\rho_{T}=\frac{1}{3a^{3}}\left(1+a^{3n}\right)\left[9+\frac{6e^{\widetilde{C}}}{1+a}+\frac{12e^{\widetilde{C}}}{1-a+a^{2}}-\frac{6ae^{\widetilde{C}}}{1-a+a^{2}}+\frac{e^{2\widetilde{C}}}{\left(1+a\right)^{2}}+\frac{2e^{2\widetilde{C}}}{\left(1+a\right)}+\frac{3e^{2\widetilde{C}}}{\left(1-a+a^{2}\right)^{2}}\right.$$
\begin{equation}\label{collapse3.40}
\left.-\frac{3ae^{2\widetilde{C}}}{\left(1-a+a^{2}\right)^{2}}+\frac{3e^{2\widetilde{C}}}{1-a+a^{2}}-\frac{2ae^{2\widetilde{C}}}{1-a+a^{2}}+9C'\right]^{\frac{1}{2}}
\end{equation}
where $\widetilde{C}$ is the integration constant. The above
relation is obtained by providing particular values to some of the
variables involved in the equation, since the differential
equation obtained was not solvable directly. Here the particular
values used are
$\omega=-\frac{1}{2}$,~$k=1$,~$n=1$~and~$\alpha=1$.~Using the
relation $\rho_{T}=\rho_{M}+\rho_{GCCG}$, we get,
\begin{equation}\label{collapse3.41}
\rho_{GCCG}=\frac{ka^{3n}\rho_{T}}{1+ka^{3n}}~~~~~~
~~~~~~~~~~~~~~~~~~\rho_{M}=\frac{\rho_{T}}{1+ka^{3n}}
\end{equation}
The expression for the time derivative of scale factor is given
by,
$$\dot{a}=-a\left[\left(\frac{1}{9a^{3}}\left(1+a^{3n}\right)\left[9+\frac{6e^{\widetilde{C}}}{1+a}+\frac{12e^{\widetilde{C}}}{1-a+a^{2}}-\frac{6ae^{\widetilde{C}}}{1-a+a^{2}}+\frac{e^{2\widetilde{C}}}{\left(1+a\right)^{2}}+\frac{2e^{2\widetilde{C}}}{\left(1+a\right)}+\frac{3e^{2\widetilde{C}}}{\left(1-a+a^{2}\right)^{2}}\right.\right.\right.$$
\begin{equation}\label{collapse3.42}
\left.\left.\left.-\frac{3ae^{2\widetilde{C}}}{\left(1-a+a^{2}\right)^{2}}+\frac{3e^{2\widetilde{C}}}{1-a+a^{2}}-\frac{2ae^{2\widetilde{C}}}{1-a+a^{2}}+9C'\right]^{\frac{1}{2}}+\frac{1}{4r_{c}^{2}}\right)^{\frac{1}{2}}+\frac{\epsilon}{2r_{c}}\right]
\end{equation}
The corresponding expressions for other physical parameters are,

$$\dot{R}(T)=-ra\left[\left(\frac{1}{9a^{3}}\left(1+a^{3n}\right)\left[9+\frac{6e^{\widetilde{C}}}{1+a}+\frac{12e^{\widetilde{C}}}{1-a+a^{2}}-\frac{6ae^{\widetilde{C}}}{1-a+a^{2}}+\frac{e^{2\widetilde{C}}}{\left(1+a\right)^{2}}+\frac{2e^{2\widetilde{C}}}{\left(1+a\right)}+\frac{3e^{2\widetilde{C}}}{\left(1-a+a^{2}\right)^{2}}\right.\right.\right.$$
\begin{equation}\label{collapse3.43}
\left.\left.\left.-\frac{3ae^{2\widetilde{C}}}{\left(1-a+a^{2}\right)^{2}}+\frac{3e^{2\widetilde{C}}}{1-a+a^{2}}-\frac{2ae^{2\widetilde{C}}}{1-a+a^{2}}+9C'\right]^{\frac{1}{2}}+\frac{1}{4r_{c}^{2}}\right)^{\frac{1}{2}}+\frac{\epsilon}{2r_{c}}\right]
\end{equation}

~~~~~and~~~~~~~~~~~~~~~~~

$$M(T)=\frac{1}{2}a^{3}r^{3}\left[\left(\frac{1}{9a^{3}}\left(1+a^{3n}\right)\left[9+\frac{6e^{\widetilde{C}}}{1+a}+\frac{12e^{\widetilde{C}}}{1-a+a^{2}}-\frac{6ae^{\widetilde{C}}}{1-a+a^{2}}+\frac{e^{2\widetilde{C}}}{\left(1+a\right)^{2}}+\frac{2e^{2\widetilde{C}}}{\left(1+a\right)}+\frac{3e^{2\widetilde{C}}}{\left(1-a+a^{2}\right)^{2}}\right.\right.\right.$$
\begin{equation}\label{collapse3.44}
\left.\left.\left.-\frac{3ae^{2\widetilde{C}}}{\left(1-a+a^{2}\right)^{2}}+\frac{3e^{2\widetilde{C}}}{1-a+a^{2}}-\frac{2ae^{2\widetilde{C}}}{1-a+a^{2}}+9C'\right]^{\frac{1}{2}}+\frac{1}{4r_{c}^{2}}\right)^{\frac{1}{2}}+\frac{\epsilon}{2r_{c}}\right]^{2}
\end{equation}.

Using the conservation equations \ref{collapse3.5} and
\ref{collapse3.6}, and using the expressions for~ $\rho_{M}$,
~$\rho_{GCCG}$~ and~$\rho_{T}$~ we get the following expression
for the interaction,
$$Q=-\frac{3\rho_{T}}{1+ka^{3n}}\left[\sqrt{\frac{\rho_{T}}{3}+\frac{1}{4r_{c}^{2}}}+\frac{\epsilon}{2r_{c}}\right]\left[\frac{1}{1+ka^{3n}}\left(ka^{3n}\left(n+1\right)+1-k^{-\alpha}\left(\frac{\rho_{T}}{1+ka^{3n}}\right)^{-(\alpha+1)}a^{-3n\alpha}\right.\right.$$
\begin{equation}\label{collapse3.45}
\left.\left.\left(C'+\left(\left(\frac{ka^{3n}\rho_{T}}{1+ka^{3n}}\right)^{1+\alpha}-C'\right)^{-\omega}\right)\right)-1\right]
\end{equation}

Fig.11, 12, 13 and 14 are for GCCG case. These are quite similar
with figs.7,8,9 and 10 respectively.

\section{Gravitational Collapse in Loop Quantum Cosmology}\label{chap04}
In recent years, loop Quantum Gravity (LQG) is an outstanding
effort to describe the quantum effect of our universe
\cite{Rovelli1, Ashtekar1}. LQG is a theory trying to quantize the
gravity with a non-perturbative and background independent method.
The theory and principles of LQG when applied in the cosmological
framework creates a new theoretical framework of Loop Quantum
Cosmology(LQC) \cite{Ashtekar2, Bojowald1, Ashtekar3}. In this
theory, classical space-time continuum is replaced by a discrete
quantum geometry. The effect of LQG can be described by the
modification of Friedmann equation to add a term quadratic in
density. In LQC, the non-perturbative effects lead to correction
term $\frac{\rho_{T}^{2}}{\rho_{1}}$ to the standard Friedmann
equation. With the inclusion of this term, the universe bounces
quantum mechanically as the matter energy density reaches the
level of $\rho_{1}$(order of Plank density).

Recently the model of DE has been explored in the framework of
LQC. The cosmological evolution in LQC has been widely studied for
quintessence and phantom DE models \cite{Wu2}. The modified
Friedmann equation for Loop Quantum Cosmology is given by
(\cite{Wu1,Chen1,Fu1})
\begin{equation}\label{collapse3.46}
H^2=\frac{\rho_{T}}{3}\left(1-\frac{\rho_{T}}{\rho_{1}}\right)
\end{equation}
Here $\rho_{1}=\sqrt{3}\pi^{2}\gamma^{3}G^{2}\hbar$ is the
critical loop quantum density and $\gamma$ is the dimensionless
Barbero-Immirzi parameter. $\rho_{T}=\rho_{M}+\rho_{E}$ represents
the total cosmic energy density, which is a sum of energy density
of DM (${\rho}_{M}$) and the energy density of DE ($\rho_{E}$).

Like the previous section in this section also we will study the
role of DM and DE separately during the collapse. Then we will
study the effect of the combination of DE and DM in the collapsing
process, both in the presence and in the absence of interaction.

\subsection{Collapse with Dark Matter}
\begin{figure}
\includegraphics[height=2in]{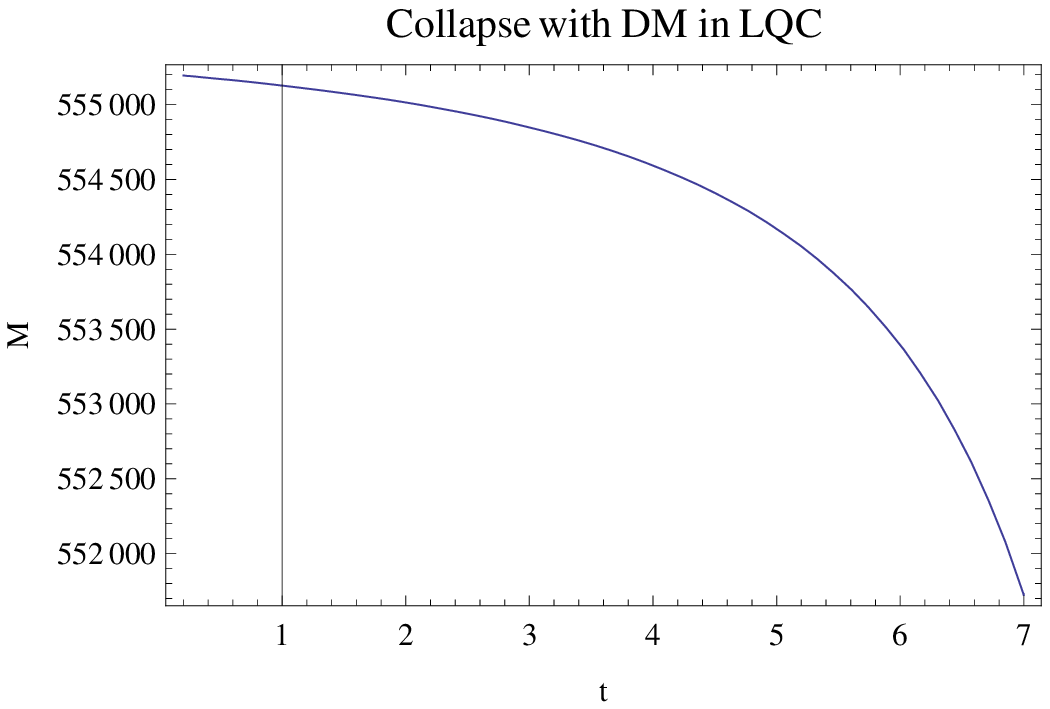}~~~~~~~~\includegraphics[height=2in]{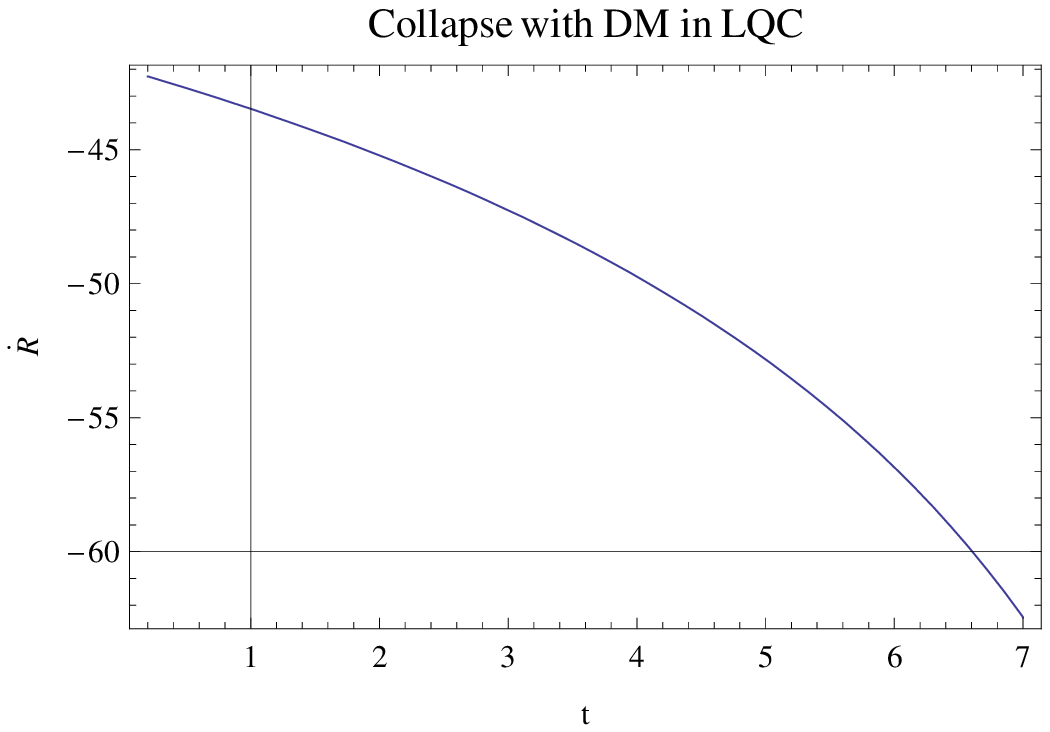}~~\\
\vspace{1mm}
~~~~~~~~~~~Fig. 15~~~~~~~~~~~~~~~~~~~~~~~~~~~~~~~~~~~~~~~~~~~~~~~~~~~~~~~~~~~~~~~~~~Fig. 16~~~~~~~~~~~~~~\\
\vspace{1mm}
\\Fig 15 : The time derivative of the radius is plotted against
time. $R_{0}=100,~ n_{0}=0.4,~N_{0}=0.1,~T=10 $ is considered.
\\Fig 16 : The mass of the collapsing cloud is plotted against
time. $R_{0}=100,~ n_{0}=0.4,~N_{0}=0.1,~T=10 $ is considered.
\end{figure}

Here $\rho_{M}\ne 0$ and $\rho_{E}=0$. From the conservation
equations (\ref{collapse3.5}) and (\ref{collapse3.6}) we get the
solution in (\ref{collapse3.14}). Putting the obtained value of
$\rho_{M}$ in equation (\ref{collapse3.46}) we get
\begin{equation}\label{collapse3.47}
a^{3}=\left[n_{0}\left(T_{0}-T\right)\right]^{2}+N_{0}^{2}
\end{equation}
where $n_{0}=\sqrt{\frac{3}{4}C_{0}}$ and
$N_{0}=\sqrt{\frac{C_{0}}{3\rho_{1}}}$ and $T_{0}$ is the
integration constant. The time derivative of the geometrical
radius of the collapsing object is determined as follows
\begin{equation}\label{collapse3.48}
\dot{R}(T)=-\frac{2}{3}R_{0}\frac{\left(a^{3}-N_{0}^{2}\right)^{\frac{1}{2}}}{a^{2}}
\end{equation}
The total mass of the collapsing cloud as determined is given by
\begin{equation}\label{collapse3.49}
M(T)=\frac{2}{9}R_{0}^{3}\frac{\left(a^{3}-N_{0}^{2}\right)}{n_{0}a^{3}}.
\end{equation}
where $R_{0}=r_{\Sigma}~n_{0}$. The limiting values of the
physical parameters are given as~
$T\rightarrow\infty,~~a\rightarrow\infty,~~\rho_{M}\rightarrow0,~~\dot{R}\rightarrow0,~~M(T)\rightarrow\frac{2R_{0}^{3}}{9n_{0}}$.
Like the previous case, the time for formation of apparent horizon
is given by the real root of the equation $\left.
\dot{R}^{2}(T_{AH})\right|_{\Sigma}
=r_{\Sigma}^{2}\dot{a}^{2}(T_{AH})=1$ as given by equation
(\ref{collapse3.11}). Thus the corresponding expression for LQC
is,
\begin{equation}\label{collapse3.50}
\left(\frac{4}{9}R_{0}^{2}\right)^{3}\left[n_{0}\left(T_{0}-T\right)\right]^{6}=\left[n_{0}^{2}\left(T_{0}-T\right)^{2}+N_{0}^{2}\right]^{4}
\end{equation}
From the above expression it can be seen that singularity will
form at the instant $T=T_{0}$.

\subsection{Collapse with Dark Energy in the form of Modified Chaplygin gas}
\begin{figure}
\includegraphics[height=2in]{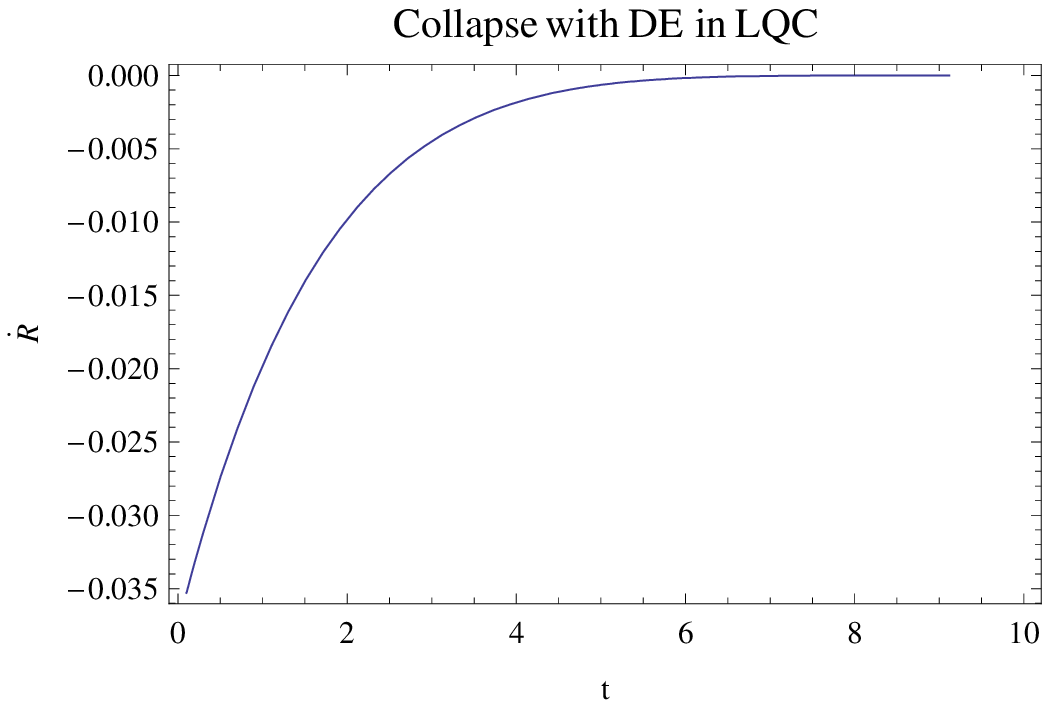}~~~~~~~~\includegraphics[height=2in]{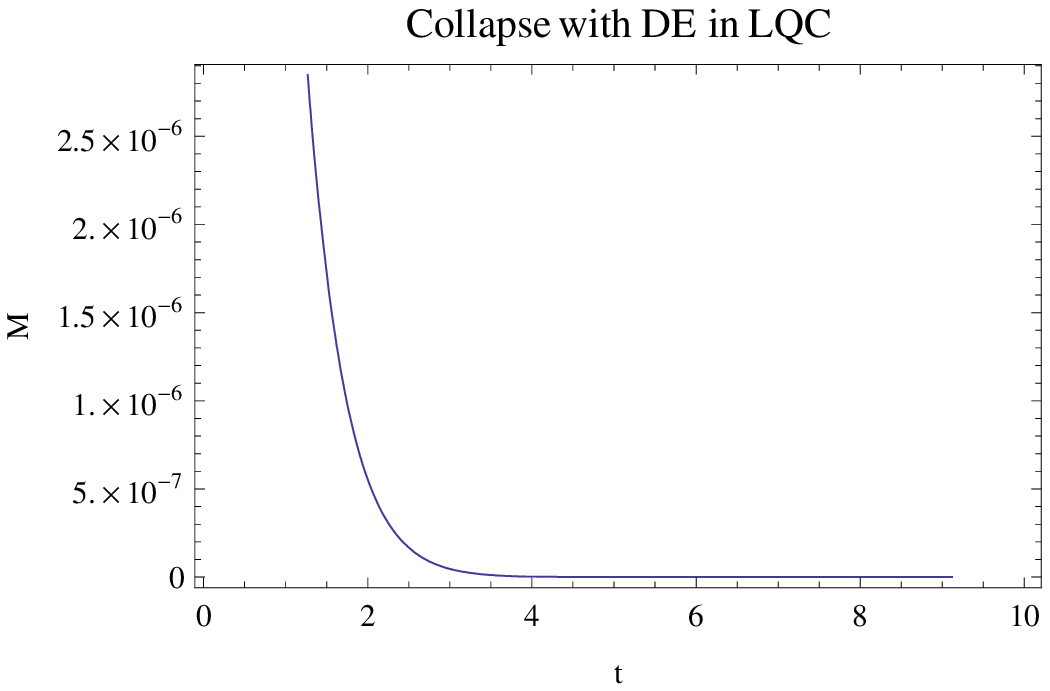}~~\\
\vspace{1mm}
~~~~~~~~~~~Fig. 17~~~~~~~~~~~~~~~~~~~~~~~~~~~~~~~~~~~~~~~~~~~~~~~~~~~~~~~~~~~~~~~~~~Fig. 18~~~~~~~~~~~~~~\\
\vspace{1mm}
\\Fig 17 : The time derivative of the radius is plotted against
time. $
r=0.5,~\rho_{1}=1000,~C_{1}=0.5,~X_{1}=0.4,~X_{2}=0.4,~X_{3}=0.4 $
is considered.
\\Fig 18 : The mass of the collapsing cloud is plotted against
time. $
r=0.5,~\rho_{1}=1000,~C_{1}=0.5,~X_{1}=0.4,~X_{2}=0.4,~X_{3}=0.4 $
is considered.
\end{figure}

In this case, $\rho_{M}=0$. The expression of the density of MCG
as given in equation (\ref{collapse3.20}) is~
$\rho_{MCG}=\left[\frac{B}{1+A}+\frac{C_{1}}{a^{3(1+A)(1+\alpha)}
} \right]^{\frac{1}{1+\alpha}}$~~ . Using this relation in the
modified Friedman equation for Loop Quantum Cosmology given in
equation (\ref{collapse3.46}), we get the expressions for the
other relevant physical quantities as follows,

\begin{equation}\label{collapse3.51}
\dot{R}(T)=-\frac{r_{\Sigma}~a}{\sqrt{3}}~\left[X_{1}+\frac{C_{1}}{a^{X_{2}}}
\right]^{X_{3}}
\left[1-\frac{1}{\rho_{1}}\left\{X_{1}+\frac{C_{1}}{a^{X_{2}}}
\right\}^{2X_{3}} \right]^{\frac{1}{2}}
\end{equation}
The total mass of the collapsing cloud as determined is given by
\begin{equation}\label{collapse3.52}
M(T)=\frac{a^{3}r_{\Sigma}^{3}}{6}
\left[X_{1}+\frac{C_{1}}{a^{X_{2}}} \right]^{X_{3}}
\left[1-\frac{1}{\rho_{1}}\left\{X_{1}+\frac{C_{1}}{a^{X_{2}}}
\right\}^{2X_{3}} \right]
\end{equation}
We know that for the cloud to undergo collapse, we should have
$\dot{R}<0$, and hence from the above expression for $\dot{R}$,we
have
\begin{equation}\label{collapse3.53}
a>\frac{C_{1}^{\frac{1}{X_{2}}}}{\left(\rho_{1}^{\frac{1}{2X_{3}}}-X_{1}\right)^\frac{1}{X_{2}}}
\end{equation}
for~~$X_{2}>0$, i.e., ~~$1+A>0$.~~~Again we should have,
\begin{equation}\label{collapse3.54}
a<\left[\frac{1}{C_{1}}\left(\rho_{1}^{\frac{1}{2X_{3}}}-X_{1}\right)\right]^{\frac{1}{\mu}}
\end{equation}
for~~$X_{2}<0$, i.e., ~~$1+A<0$, where $X_{2}=-\mu$. The limiting
values for the physical parameters are given as follows:
$$\textbf{case1}$$
$$~~~~~a\rightarrow0:~~~~\rho_{MCG}\rightarrow\infty,~~~~for ~~~1+A>0;~~~~\rho_{MCG}\rightarrow\left[\frac{B}{1+A}\right]^{\frac{1}{1+\alpha}},~~~~for~~~~1+A<0$$
$$~~~~\dot{R}(T)\rightarrow-\frac{r_{\Sigma}}{\sqrt{3\rho_{1}}}a^{-(2+3A)}C_{1}^{2X_{3}},~~~~for~~1+A<0,~~i.e.,~~a<\left[\frac{1}{C_{1}}\left(\rho_{1}^{\frac{1}{2X_{3}}}-X_{1}\right)\right]^{\frac{1}{\mu}}$$
$$~~~~M(T)\rightarrow\frac{r_{\Sigma}^{3}}{6\rho_{1}}C_{1}^{4X_{3}}a^{-3(1+2A)},~~for~~1+A<0$$

$$\textbf{case2}$$
$$~~~~a\rightarrow\infty:~~~~\rho_{MCG}\rightarrow\left[\frac{B}{1+A}\right]^{\frac{1}{1+\alpha}},~~~~for ~~~1+A>0;~~~~\rho_{MCG}\rightarrow\infty,~~~~for ~~~1+A<0 $$
$$~~~~\dot{R}(T)\rightarrow-\frac{r_{\Sigma}}{\sqrt{3\rho_{1}}}a^{-(2+3A)}C_{1}^{2X_{3}},~~~~for~~1+A>0,~~i.e.,~~a>\frac{C_{1}^{\frac{1}{X_{2}}}}{\left(\rho_{1}^{\frac{1}{2X_{3}}}-X_{1}\right)^\frac{1}{X_{2}}}$$
$$~~~~M(T)\rightarrow\frac{r_{\Sigma}^{3}}{6\rho_{1}}C_{1}^{4X_{3}}a^{-3(1+2A)},~~for~~1+A>0$$
The cloud will start to be trapped at the instant given by the
real roots of the equation,
\begin{equation}\label{collapse3.55}
\frac{r_{\Sigma}^{2}~a^{2}}{3}~\left[X_{1}+\frac{C_{1}}{a^{X_{2}}}\right]^{2X_{3}}\left[1-\frac{1}{\rho_{1}}\left\{X_{1}+\frac{C_{1}}{a^{X_{2}}}\right\}^{2X_{3}}\right]=1
\end{equation}

\subsection{Collapse with Dark Energy in the form of Generalised Cosmic Chaplygin Gas}
\begin{figure}
\includegraphics[height=2in]{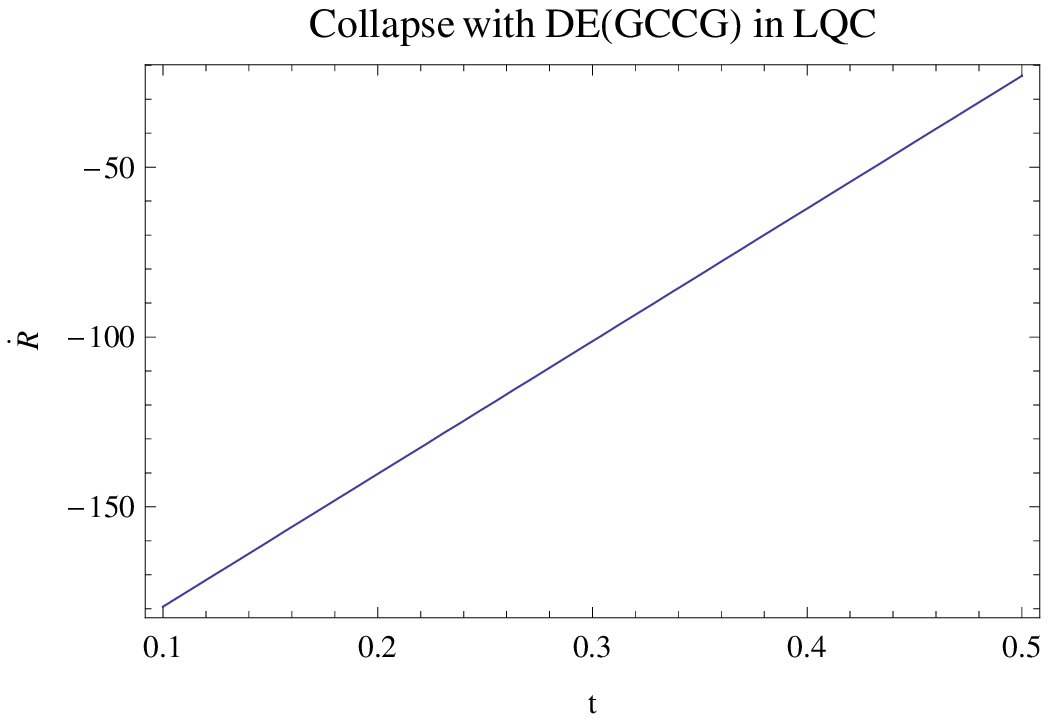}~~~~~~~~\includegraphics[height=2in]{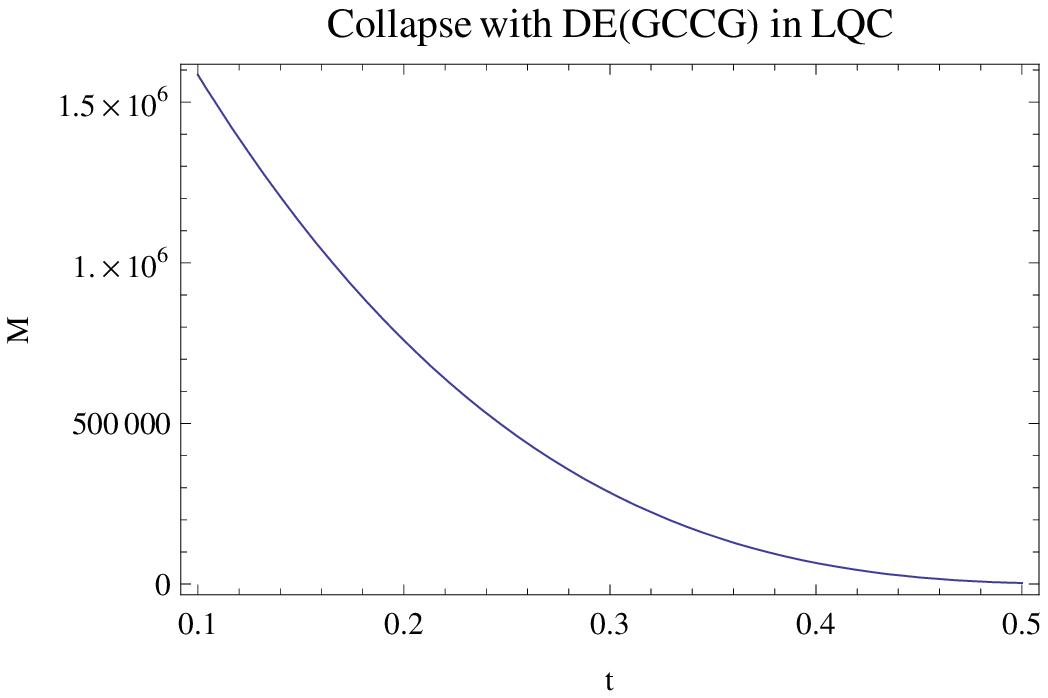}~~\\
\vspace{1mm}
~~~~~~~~~~~Fig. 19~~~~~~~~~~~~~~~~~~~~~~~~~~~~~~~~~~~~~~~~~~~~~~~~~~~~~~~~~~~~~~~~~~Fig. 20~~~~~~~~~~~~~~\\
\vspace{1mm}
\\Fig 19 : The time derivative of the radius is plotted against
time. $
r=1000,~\rho_{1}=1000,~B'=5,~C'=100,~X_{1}=0.00005,~X_{2}=1.5,~X_{3}=0.25
$ is considered.
\\Fig 20 : The mass of the collapsing cloud is plotted against
time. $
r=1000,~\rho_{1}=1000,~B'=5,~C'=100,~X_{1}=0.00005,~X_{2}=1.5,~X_{3}=0.25$
is considered.
\end{figure}

Here $\rho_{M}=0$. The expression for the density of DE is given
by equation (\ref{collapse3.24}). The expressions for the relevant
physical quantities are,
\begin{equation}\label{collapse3.56}
\dot{R}(T)=-\frac{ar_{\Sigma}}{\sqrt{3}}\left[C'+\left(1+\frac{B'}{a^{X_{2}'}}\right)^{X_{1}'}\right]^{X_{3}'}\left[1-\frac{1}{\rho_{1}}\left(C'+\left(1+\frac{B'}{a^{X_{2}'}}\right)^{X_{1}'}\right)^{2X_{3}'}\right]^{\frac{1}{2}}
\end{equation}
and
\begin{equation}\label{collapse3.57}
M(T)=\frac{a^{3}r_{\Sigma}^{3}}{6}\left[C'+\left(1+\frac{B'}{a^{X_{2}'}}\right)^{X_{1}'}\right]^{2X_{3}'}\left[1-\frac{1}{\rho_{1}}\left(C'+\left(1+\frac{B'}{a^{X_{2}'}}\right)^{X_{1}'}\right)^{2X_{3}'}\right]
\end{equation}
For the cloud to undergo collapse, we should have,
\begin{equation}\label{collapse3.58}
a^{3}>\frac{1}{\rho_{1}}\left[C'a^{X_{1}'X_{2}'}+\left(a^{X_{2}'}+B'\right)^{X_{1}'}\right]^{2X_{3}'}
\end{equation}
The limiting values of the corresponding physical parameters are
given as follows,
$$\textbf{case1}$$
$$~~~~~a\rightarrow0:~~~~\rho_{GCCG}\rightarrow\infty,~~~~for ~~~1+\omega>0;~~~~\rho_{GCCG}\rightarrow\left[C'+1\right]^{\frac{1}{1+\alpha}},~~~~for~~~~1+\omega<0$$
$$~~~~\dot{R}(T)\rightarrow-\frac{r_{\Sigma}a^{-2}}{\sqrt{3\rho_{1}}}B'^{\frac{1}{(1+\alpha)(1+\omega)}},~~~~for~~1+\omega>0$$
$$~~~~M(T)\rightarrow\frac{r_{\Sigma}^{3}}{6\sqrt{\rho_{1}}}B'^{\frac{2}{(1+\alpha)(1+\omega)}},~~for~~1+\omega>0$$

$$\textbf{case2}$$
$$~~~~a\rightarrow\infty:~~~~\rho_{GCCG}\rightarrow\infty,~~~~for ~~~1+\omega<0;~~~~\rho_{GCCG}\rightarrow\left[C'+1\right]^{\frac{1}{1+\alpha}},~~~~for~~~~1+\omega>0$$
$$~~~~\dot{R}(T)\rightarrow-\frac{r_{\Sigma}a}{\sqrt{3\rho_{1}}}\left(C'+1\right)^{\frac{1}{2(1+\alpha)}}\left[\rho_{1}-\left(C'+1\right)^{\frac{1}{1+\alpha}}\right]^{\frac{1}{2}},~~~~for~~1+\omega>0$$
$$~~~~M(T)\rightarrow\frac{r_{\Sigma}^{3}a^{6}}{6\sqrt{\rho_{1}}}\left(C'+1\right)^{\frac{1}{1+\alpha}}\left[\rho_{1}-\left(C'+1\right)^{\frac{1}{1+\alpha}}\right],~~for~~1+\omega>0$$
All the above limiting values are calculated considering that
equation (\ref{collapse3.58}) is satisfied. The time for the
formation of the apparent horizon is given by the real root of the
following equation,
\begin{equation}\label{collapse3.59}
\frac{a^{2}r_{\Sigma}^{2}}{3}\left[C'+\left(1+\frac{B'}{a^{X_{2}'}}\right)^{X_{1}'}\right]^{2X_{3}'}\left[1-\frac{1}{\rho_{1}}\left(C'+\left(1+\frac{B'}{a^{X_{2}'}}\right)^{X_{1}'}\right)^{2X_{3}'}\right]=1
\end{equation}

\subsection{Effect of combination of Dark Energy(in the form of MCG) and Dark Matter}
\subsubsection{Case I : $Q = 0$ i.e., No Interaction Between
Dark Matter And Dark Energy:}
\begin{figure}
\includegraphics[height=2in]{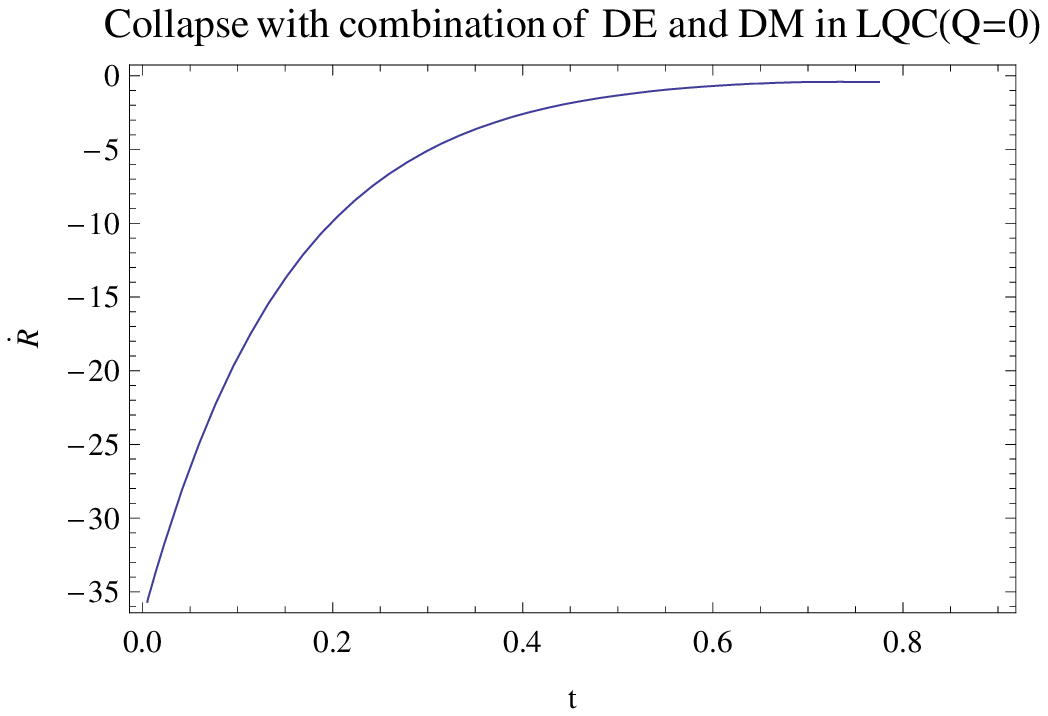}~~~~~~~~\includegraphics[height=2in]{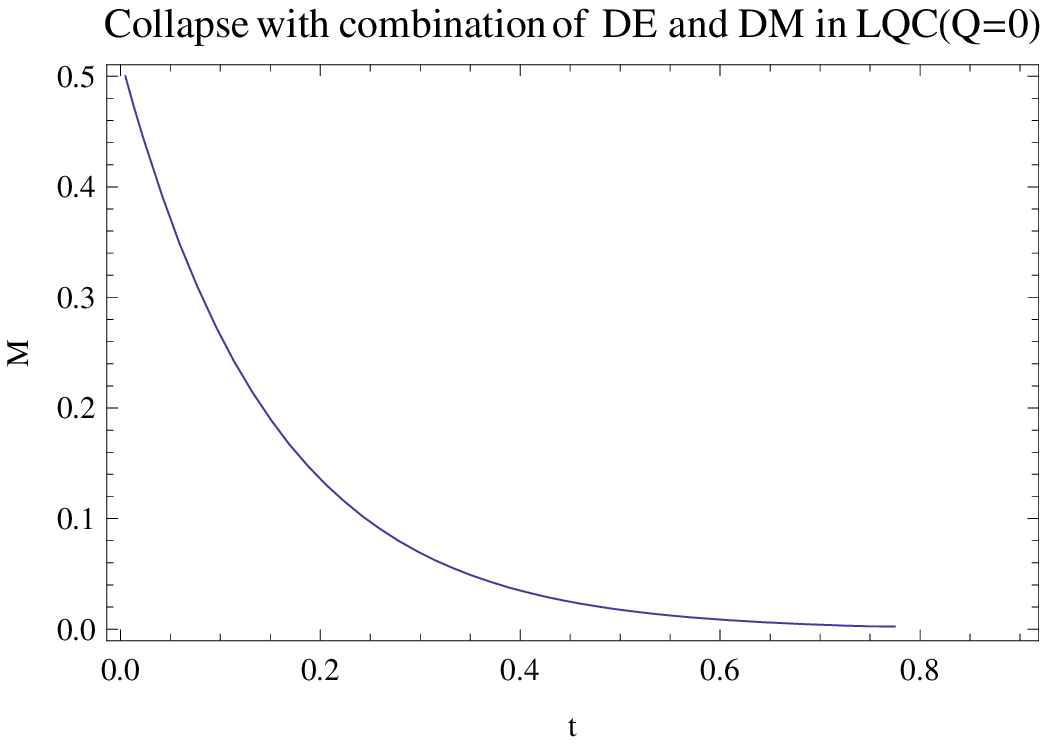}~~\\
\vspace{1mm}
~~~~~~~~~~~Fig. 21~~~~~~~~~~~~~~~~~~~~~~~~~~~~~~~~~~~~~~~~~~~~~~~~~~~~~~~~~~~~~~~~~~Fig. 22~~~~~~~~~~~~~~\\
\vspace{1mm}
\\Fig 21 : The time derivative of the radius is plotted against
time. $
r=10,~\rho_{1}=1000,~B'=5,~C_{1}=500,~C_{0}=0.00001,~X_{1}=40,~X_{2}=0.04,~X_{3}=0.4
$ is considered.
\\Fig 22 : The mass of the collapsing cloud is plotted against
time. $
r=10,~\rho_{1}=1000,~B'=5,~C_{1}=500,~C_{0}=0.00001,~X_{1}=40,~X_{2}=0.04,~X_{3}=0.4$
is considered.
\end{figure}

\begin{figure}
\includegraphics[height=2in]{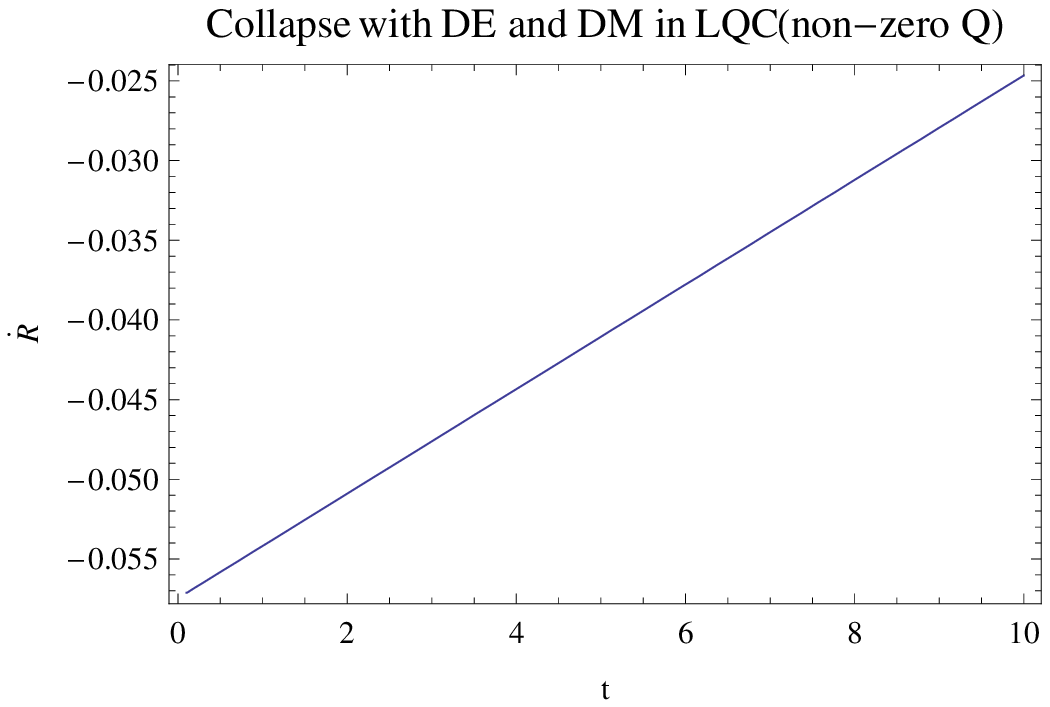}~~~~~~~~\includegraphics[height=2in]{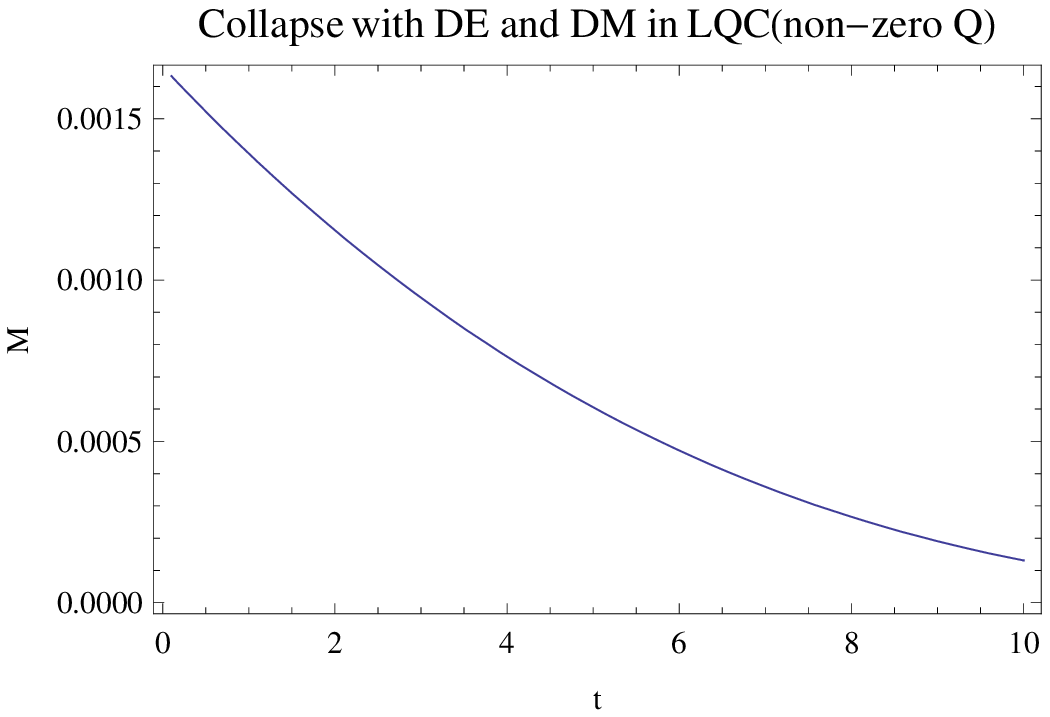}~~\\
\vspace{1mm}
~~~~~~~~~~~Fig. 23~~~~~~~~~~~~~~~~~~~~~~~~~~~~~~~~~~~~~~~~~~~~~~~~~~~~~~~~~~~~~~~~~~Fig. 24~~~~~~~~~~~~~~\\
\vspace{1mm}
\\Fig 23 : The time derivative of the radius is plotted against
time. $ r=10,~\rho_{1}=10,~B=5,~C=5,~,~n=1 $ is considered.
\\Fig 24 : The mass of the collapsing cloud is plotted against
time. $ r=10,~\rho_{1}=10,~B=5,~C=5,~,~n=1$ is considered.
\end{figure}
The solution for $\rho_{M}$ and $\rho_{MCG}$ are given in
equations (\ref{collapse3.14}) and (\ref{collapse3.20}). The
relevant physical quantities are obtained as

\begin{equation}\label{collapse3.60}
\dot{R}(T)=-\frac{r_{\Sigma}~a}{\sqrt{3}}~\left[\frac{C_{0}}{a^{3}}+
\left(X_{1}+\frac{C_{1}}{a^{X_{2}}}
\right)^{2X_{3}}\right]^{\frac{1}{2}}  \left[1-\frac{1}{\rho_{1}}
\left\{\frac{C_{0}}{a^{3}}+ \left(X_{1}+\frac{C_{1}}{a^{X_{2}}}
\right)^{2X_{3}}\right\} \right]^{\frac{1}{2}}
\end{equation}
and
\begin{equation}\label{collapse3.61}
M(T)=\frac{r_{\Sigma}^{3}~a^{3}}{6}~\left[\frac{C_{0}}{a^{3}}+
\left(X_{1}+\frac{C_{1}}{a^{X_{2}}} \right)^{2X_{3}}\right]
\left[1-\frac{1}{\rho_{1}} \left\{\frac{C_{0}}{a^{3}}+
\left(X_{1}+\frac{C_{1}}{a^{X_{2}}} \right)^{2X_{3}}\right\}
\right]
\end{equation}
In this case, the condition for the cloud to undergo collapse is
given by,
\begin{equation}\label{collapse3.62}
a^{X_{2}}\left[\left(\rho_{1}^{2}-\frac{C_{0}}{a^{3}}\right)^{\frac{1}{2X_{3}}}-X_{1}\right]>C_{1}
\end{equation}
for~~$X_{2}>0$, i.e., ~~$1+A>0$,~~~and $a\rightarrow~\infty$. The
above result is also valid for ~~$X_{2}<0,$ ~i.e., ~$1+A<0$,~~and
$a\rightarrow0$. ~~~~~The limiting values of the physical
parameters are as follows,
$$\textbf{case1}$$
$$~~~~~a\rightarrow0:~~~\rho_{M}\rightarrow\infty,~~~~~\rho_{MCG}\rightarrow\infty,~~~~for ~~~1+A>0;~~~~~\rho_{MCG}\rightarrow\left[\frac{B}{1+A}\right]^{\frac{1}{1+\alpha}},~~~~~for ~~~~1+A<0 $$
$$~~~~~\dot{R}(T)\rightarrow-\frac{r_{\Sigma}}{\sqrt{3}\rho_{1}}a^{-\frac{1}{2}(1+3A)}\left(\frac{C_{0}}{a^{3}}+X_{1}^{2X_{3}}\right)^{\frac{1}{2}}\left[\rho_{1}-\left(\frac{C_{0}}{a^{3}}+X_{1}^{2X_{3}}\right)^{\frac{1}{2}}\right],~~~~for ~~~1+A<0$$
$$~~~~M(T)\rightarrow0,~~~~for~~~~1+A>0,~~M(T)\rightarrow\infty,~~~~for~~~~1+A<0$$

 $$\textbf{case2}$$
$$~~~~~a\rightarrow\infty:~~~~~\rho_{M}\rightarrow0,~~~~\rho_{MCG}\rightarrow\left[\frac{B}{1+A}\right]^{\frac{1}{1+\alpha}},~~~~for ~~~1+A>0;~~~~~\rho_{MCG}\rightarrow\infty,~~~~for ~~~1+A<0 $$
$$~~~~~\dot{R}(T)\rightarrow-\frac{r_{\Sigma}}{\sqrt{3}\rho_{1}}a^{-\frac{1}{2}(1+3A)}\left(\frac{C_{0}}{a^{3}}+X_{1}^{2X_{3}}\right)^{\frac{1}{2}}\left[\rho_{1}-\left(\frac{C_{0}}{a^{3}}+X_{1}^{2X_{3}}\right)^{\frac{1}{2}}\right],~~~~for ~~~1+A>0$$
$$~~~~M(T)\rightarrow0,~~~~for~~~~1+A<0,~~M(T)\rightarrow\infty,~~~~for~~~~1+A>0$$

\subsubsection{Case II : $Q \neq 0$ i.e., Interaction Between
Dark Matter And Dark Energy:}
As in the case of DGP brane here also we will assume the relation
$\frac{\rho_{MCG}}{\rho_{M}}=Ca^{3n}$ as given in equation
(\ref{collapse3.30}). The expressions for $\rho_{M}$, $\rho_{MCG}$
and $\rho_{T}$ are given in equations (\ref{collapse3.14}),~
(\ref{collapse3.24}) and (\ref{collapse3.31}) respectively. The
expression for interaction is given by
\begin{equation}\label{collapse3.63}
Q=-\frac{3\rho_{T}}{1+Ca^{3n}}\left[\sqrt{\frac{\rho_{T}}{3}\left(1-\frac{\rho_{T}}{\rho_{1}}\right)}\right]\left[\frac{1}{1+Ca^{3n}}\left(1+a^{3n}\left(C\left(1+A+3n\right)\right)-BC^{-\alpha}a^{-3\alpha
n}\left(1+Aa^{3n}\right)^{\alpha+1}\rho_{T}^{-\alpha-1}\right)-1\right]
\end{equation}
In this case the value for the gradient of scale factor is given
by
\begin{equation}\label{collapse3.64}
\dot{a}=-a\left[\sqrt{\frac{\rho_{T}}{3}\left(1-\frac{\rho_{T}}{\rho_{1}}\right)}\right]
\end{equation}
where $\rho_{T}$ is given by (\ref{collapse3.31}). The
corresponding expressions for mass $M(T)$ and $\dot{R}$ are given
as follows:
\begin{equation}\label{collapse3.65}
\dot{R}(T)=-r_{\Sigma}a\sqrt{\frac{\rho_{T}}{3}\left(1-\frac{\rho_{T}}{\rho_{1}}\right)}
\end{equation}
and
\begin{equation}\label{collapse3.66}
M(T)=\frac{1}{6}a^{3}r_{\Sigma}^{3}\rho_{T}^{2}\left(1-\frac{\rho_{T}}{\rho_{1}}\right)
\end{equation}
The limiting values for the physical parameters for this case are
given by
$$\textbf{case1}$$
$$ a\rightarrow0:~~~~\rho_{M}\rightarrow~a^{-3n},~~~~\rho_{MCG}\rightarrow~a~ constant,~~~~\rho_{T}\rightarrow~a^{-3n},~~~~\dot{R}\rightarrow~-a^{1-\frac{3n}{2}},~~~~M(T)\rightarrow~a^{3(1-n)}$$

$$\textbf{case2}$$
$$ a\rightarrow~\infty:~~~~\rho_{M}\rightarrow~a^{-3n(\alpha+1)(A+1)},~~~~\rho_{MCG}\rightarrow~a^{-3n(\alpha+1)(A+1)},~~~~\rho_{T}\rightarrow~a^{-3n(\alpha+1)(A+1)},~~~~\dot{R}\rightarrow~-a^{1-\frac{3n}{2}(1+A)}$$
$$M(T)\rightarrow~a^{3[1-n(1+A)]}$$

\subsection{Effect of combination of Dark Energy(in the form of GCCG) and Dark Matter}
\subsubsection{Case I : $Q = 0$ i.e., Non-Interaction Between
Dark Matter And Dark Energy:}
\begin{figure}

\includegraphics[height=2in]{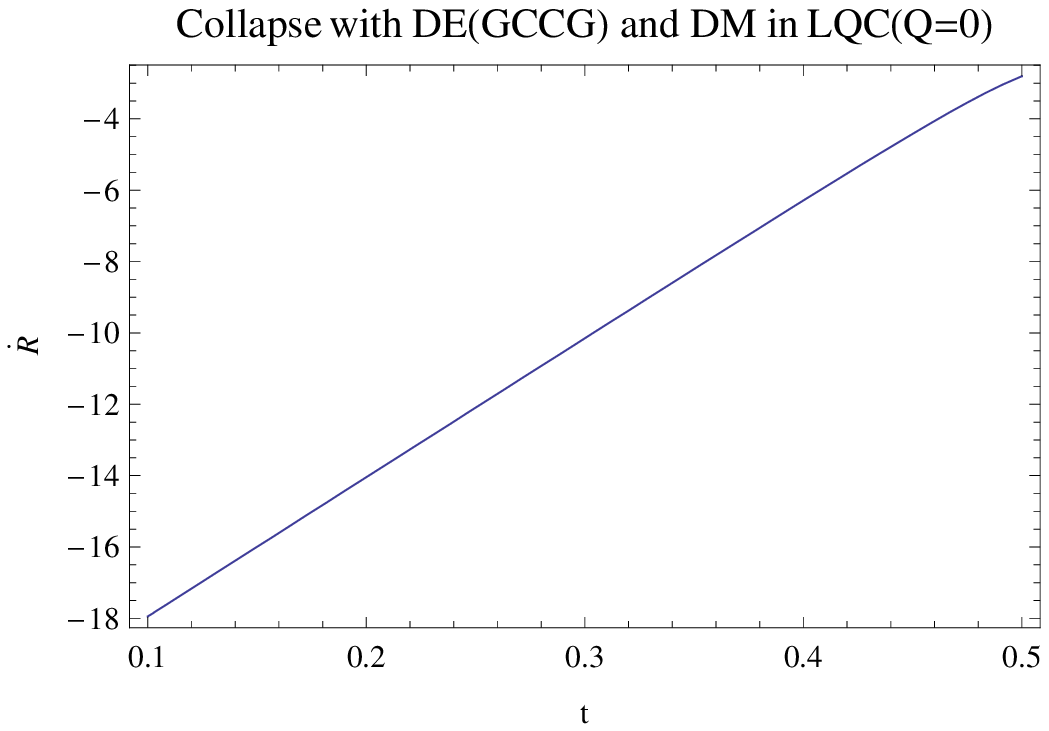}~~~~~~~~\includegraphics[height=2in]{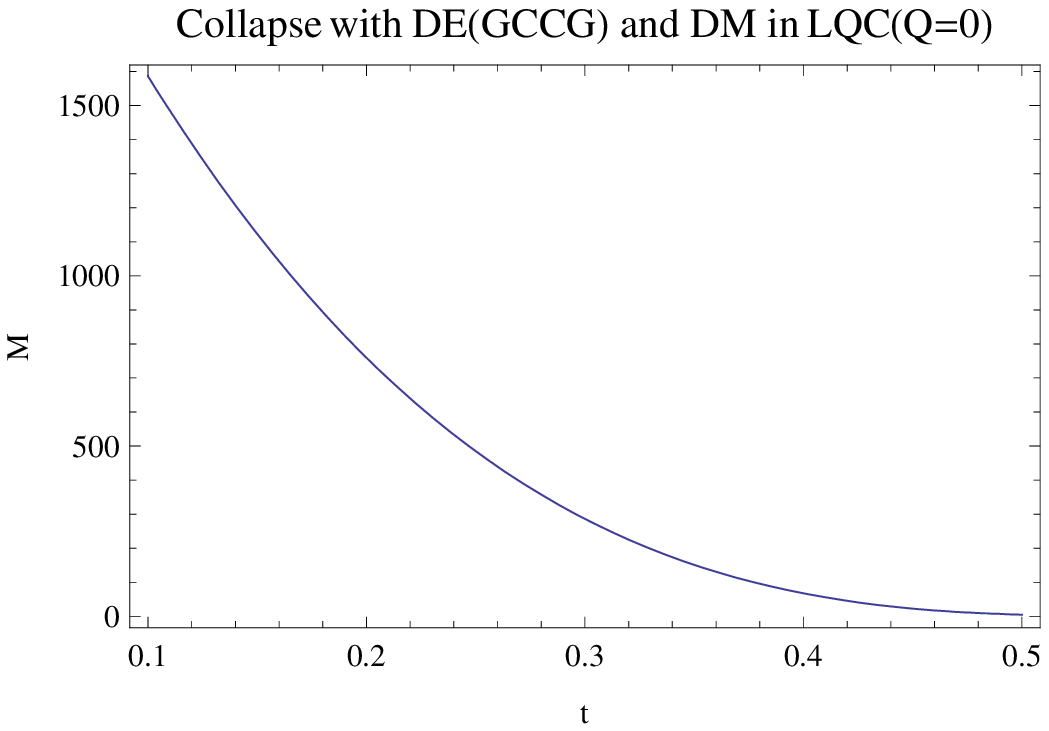}~~\\
\vspace{1mm}
~~~~~~~~~~~Fig. 25~~~~~~~~~~~~~~~~~~~~~~~~~~~~~~~~~~~~~~~~~~~~~~~~~~~~~~~~~~~~~~~~~~Fig. 26~~~~~~~~~~~~~~\\
\\Fig 25 : The time derivative of the radius is plotted against
time. $
r=100,~\rho_{1}=1000,~B'=5,~C'=100,~,~n=1,~X_{1}=0.00005,~X_{2}=1.5,~X_{3}=0.25,~C_{0}=0.00001
$ is considered.
\\Fig 26 : The mass of the collapsing cloud is plotted against time. $
r=100,~\rho_{1}=1000,~B'=5,~C'=100,~,~n=1,~X_{1}=0.00005,~X_{2}=1.5,~X_{3}=0.25,~C_{0}=0.00001
$ is considered. \vspace{1mm}
\end{figure}

\begin{figure}

\includegraphics[height=2in]{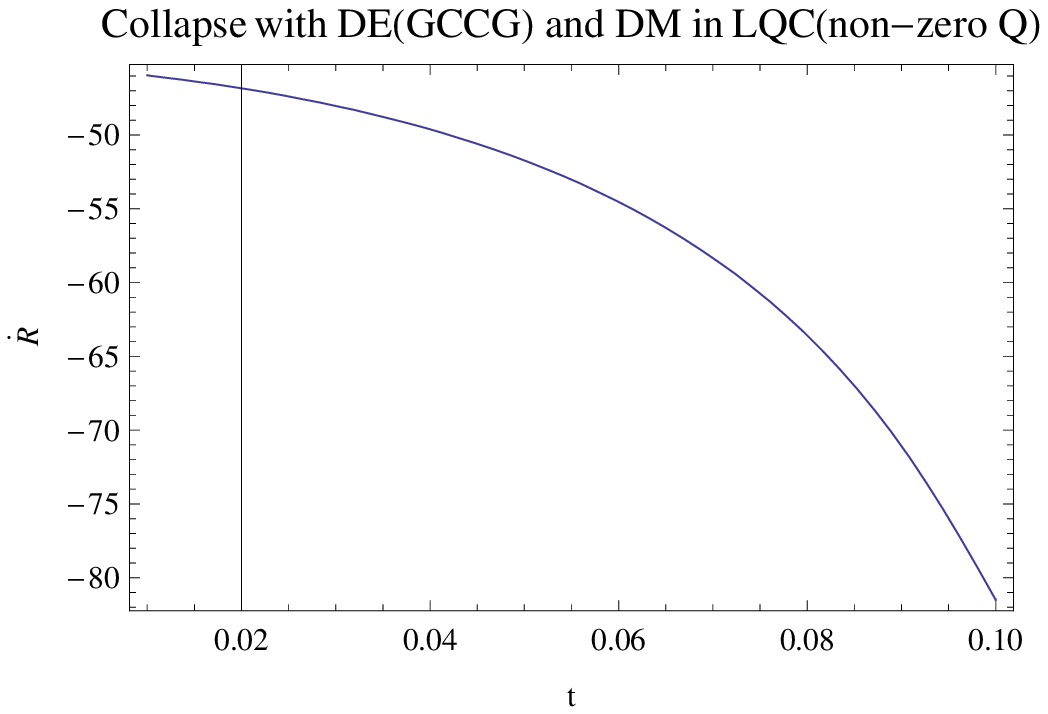}~~~~~~~~\includegraphics[height=2in]{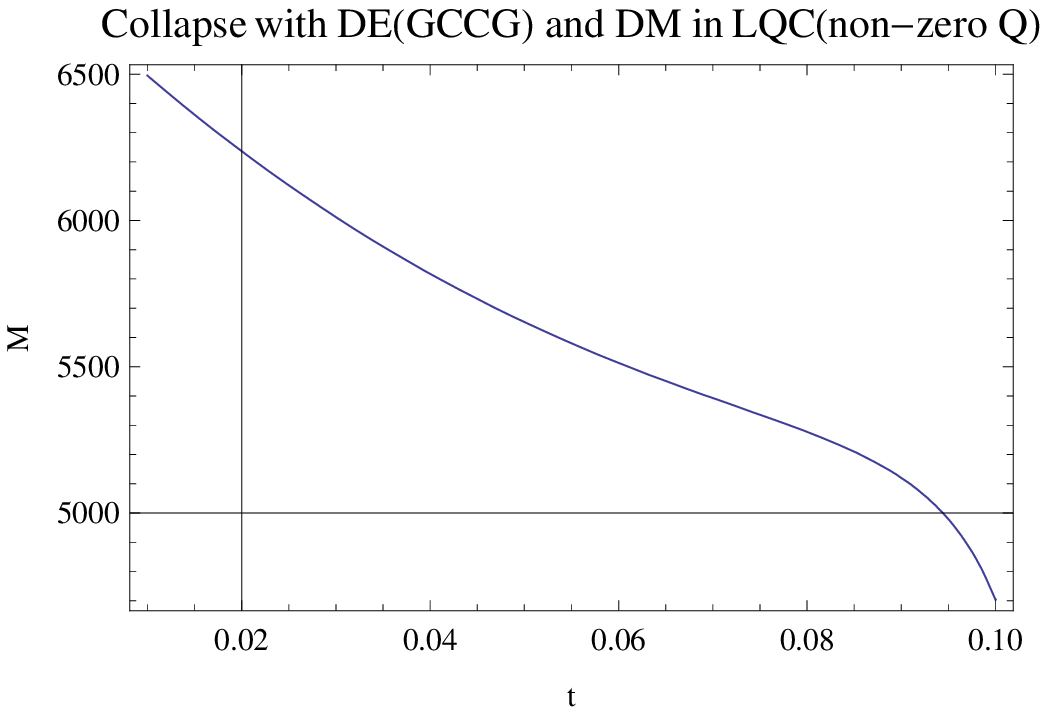}~~\\
\vspace{1mm}
~~~~~~~~~~~Fig. 27~~~~~~~~~~~~~~~~~~~~~~~~~~~~~~~~~~~~~~~~~~~~~~~~~~~~~~~~~~~~~~~~~~Fig. 28~~~~~~~~~~~~~~\\
\\Fig 27 : The time derivative of the radius is plotted against
time. $ r=10,~\rho_{1}=100000,~B'=5,~\widetilde{C}=0.001,~n=1 $ is
considered.
\\Fig 28 : The mass of the collapsing cloud is plotted against time. $
 r=10,~\rho_{1}=100000,~B'=5,~\widetilde{C}=0.001,~n=1$
is considered. \vspace{1mm}

\end{figure}

The solution for $\rho_{M}$ and $\rho_{GCCG}$ are given in
equations (\ref{collapse3.14}) and (\ref{collapse3.24})
respectively. The relevant physical quantities are obtained as
\begin{equation}\label{collapse3.67}
\dot{R}(T)=-\frac{ar_{\Sigma}}{\sqrt{3}}\left[\frac{C_{0}}{a^{3}}+\left[C'+\left(1+\frac{B'}{a^{X_{2}'}}\right)^{X_{1}'}\right]^{2X_{3}'}\right]^{\frac{1}{2}}\left[1-\frac{1}{\rho_{1}}\left[\frac{C_{0}}{a^{3}}+\left(C'+\left(1+\frac{B'}{a^{X_{2}'}}\right)^{X_{1}'}\right)^{2X_{3}'}\right]\right]^{\frac{1}{2}}
\end{equation}
and
\begin{equation}\label{collapse3.68}
M(T)=\frac{a^{3}r_{\Sigma}^{3}}{6}\left[\frac{C_{0}}{a^{3}}+\left[C'+\left(1+\frac{B'}{a^{X_{2}'}}\right)^{X_{1}'}\right]^{2X_{3}'}\right]\left[1-\frac{1}{\rho_{1}}\left[\frac{C_{0}}{a^{3}}+\left(C'+\left(1+\frac{B'}{a^{X_{2}'}}\right)^{X_{1}'}\right)^{2X_{3}'}\right]\right]
\end{equation}
The collapsing process will take place provided $\dot{R}(T)<0$,
hence the corresponding condition is given by,
\begin{equation}\label{collapse3.69}
a^{3}>\frac{1}{\rho_{1}}\left[C_{0}+\left(C'a^{X{1}'X_{2}'}+\left(a^{X_{2}'}+B'\right)^{X_{1}'}\right)^{2X_{3}'}\right]
\end{equation}
The corresponding limiting values of different parameters are
given below,
$$\textbf{case1}$$
$$~~~~~a\rightarrow0:~~~~\rho_{GCCG}\rightarrow\infty,~~~~for ~~~1+\omega>0;~~~~\rho_{GCCG}\rightarrow\left[C'+1\right]^{\frac{1}{1+\alpha}},~~~~for~~~~1+\omega<0$$
$$~~~~\dot{R}(T)\rightarrow-\frac{r_{\Sigma}a^{-2}}{\sqrt{3\rho_{1}}}\left(C_{0}+B'^{\frac{1}{(1+\alpha)(1+\omega)}}\right),~~~~for~~1+\omega>0$$
$$~~~~M(T)\rightarrow\frac{r_{\Sigma}^{3}}{6\sqrt{\rho_{1}}}\left(C_{0}+B'^{\frac{1}{1+\omega}}\right)^{1+\alpha},~~for~~1+\omega>0$$

$$\textbf{case2}$$
$$~~~~a\rightarrow\infty:~~~~\rho_{GCCG}\rightarrow\infty,~~~~for ~~~1+\omega<0;~~~~\rho_{GCCG}\rightarrow\left[C'+1\right]^{\frac{1}{1+\alpha}},~~~~for~~~~1+\omega>0$$
$$~~~~\dot{R}(T)\rightarrow-\frac{r_{\Sigma}a}{\sqrt{3\rho_{1}}}\left(C'+1\right)^{\frac{1}{2(1+\alpha)}}\left[\rho_{1}-\left(C'+1\right)^{\frac{1}{1+\alpha}}\right]^{\frac{1}{2}},~~~~for~~1+\omega>0$$
$$~~~~M(T)\rightarrow\frac{r_{\Sigma}^{3}a^{6}}{6\sqrt{\rho_{1}}}\left(C'+1\right)^{\frac{1}{1+\alpha}}\left[\rho_{1}-\left(C'+1\right)^{\frac{1}{1+\alpha}}\right],~~for~~1+\omega>0$$

\subsubsection{Case II : $Q \neq 0$ i.e., Interaction Between
Dark Matter And Dark Energy:}
As usual we consider $\frac{\rho_{GCCG}}{\rho_{M}}=ka^{3n}$, where
$k>0$ and $'n'$ is an arbitrary constant. Just like the previous
cases the expressions for $\rho_{M}$,~ $\rho_{GCCG}$ and
~$\rho_{T}$~ are also taken into account from equations~
(\ref{collapse3.14})~,~(\ref{collapse3.24})~ and
(\ref{collapse3.40}) respectively. The relevant parameters, i.e,
~$\dot{a}$,~ $\dot{R}(T)$ and $M(T)$ are given by the equations
(\ref{collapse3.64}),~(\ref{collapse3.65})~ and
(\ref{collapse3.66}) respectively using the values of $\rho_{M}$,~
$\rho_{GCCG}$ and ~$\rho_{T}$~ from the equations~
(\ref{collapse3.14})~,~(\ref{collapse3.24})~ and
(\ref{collapse3.40}) respectively.

Using the conservation equations (5) and (6), we calculate the
interaction as given below,

$$Q=-\frac{3\rho_{T}}{1+ka^{3n}}\sqrt{\frac{\rho_{T}}{3}\left(1-\frac{\rho_{T}}{\rho_{1}}\right)}\left[\frac{1}{1+ka^{3n}}\left(ka^{3n}\left(n+1\right)+1-k^{-\alpha}\left(\frac{\rho_{T}}{1+ka^{3n}}\right)^{-(\alpha+1)}a^{-3n\alpha}\right.\right.$$
\begin{equation}\label{collapse3.70}
\left.\left.\left(C'+\left(\left(\frac{ka^{3n}\rho_{T}}{1+ka^{3n}}\right)^{1+\alpha}-C'\right)^{-\omega}\right)\right)-1\right]
\end{equation}

 From figs. 15 to 28 we have
demonstrated the $\dot{R}$ and mass curves for LQC. Here the
$\dot{R}$ for DM collapse is increasing in magnitude with time.
But the mass is decreasing. The fact that the velocity of inward
collapse is increasing but mass is decreasing, is quite an awkward
incident to explain. This is a very unique property for LQC. In
figs.17 and 18, we see the DE collapse in the universe filled up
with MCG. The inward velocity with time whereas the mass of the
collapsing centre decreases. The same result is followed for GCCG
case in graphs figs.19 and 20. Unlike DGP case in LQC DM and DE
combined case is dominated by the speedy collapse and decreasing
mass incident.

\section{Detailed Graphical Analysis}

In the above figures the time derivative of the geometrical
radius($\dot{R}$) of the collapsing object and the mass($M$) of
the same have been plotted against time, for all the above cases.
From the plots it is evident that the time derivative of the
radius inevitably takes negative values, which in turn shows that
the radius decreases with time, thus resulting in gravitational
collapse. We discuss the graphs for the two sections separately in
detail.

~~~~~~~~~~~~~~~~~~~~~~~~~~~~~~~~~~~~~~~~~~~~~~~~~~~~~~~~~~~~~~~~~~~~~~~~~~~~~~~~~~~~~~~~~~~~~~~~~~~~~~~~~~~~~~~~~~~~~~~~~~~~~~~~~~~~~~~~~~~~~~~~~~~~~~~~~~~~~~~~~~~~~~~~~~~~~~~~~~~~~~~~~~~~~~~~~~~~~~~~~~~~~~~~~~~~~~~~~~~~~~~~~~~~~~~~~~~~~~~~~~~~~~~~~~~~~~~~~~~~~~~~~~~~~~~~~~~~~~~~~~~~~~~~~~~~~~~~~~~~~~~~~~~~~~~~~~~~~~~~~~~~~
~~~~~~~~~~~~~~~~~~~~~~~~~~~~~~~~~~~~~~~~~~~~~~~~~~~~~~~~~~~~~~~~~~~

~~~~~~~~~~~~~~~~~~~~~~~~~~~~~~\textbf{DGP BRANE MODEL}~~~~~~~~~~

From the first figure we see that there is a considerable decrease
in the value of $\dot{R}$, towards more negative values in the
presence of dark matter(dust) only. This shows that matter field
collapses due to their inward gravitational pull, and gradually
increases in mass by accumulating more matter, which is evident
from fig.2. From fig.3, it is seen that in the presence of dark
energy in the form of MCG, value of $\dot{R}$ increases with time
and has a gradual tendency of stepping into the positive region.
This is due to the fact that the large negative pressure of the DE
does not facilitate collapse under normal conditions. Initially if
the space-time is trapped, during evolution it gets un-trapped and
eventually the cloud shows a tendency to expand due to the
presence of DE in the form of MCG. Even if the cloud does not
expand, due to its local condensation, the time for formation of
the apparent horizon and consequently a BH will be much delayed in
future. Hence the formation of BH is quite an uncertainty, if at
all it is a possibility. Fig.4 shows that due to DE accretion on
the collapsing object, there is a gradual decrease in its mass,
with time. With GCCG as the DE the story is almost the same, as is
evident from figures 5 and 6. Fig. 7 shows that in the absence of
any interaction, a combination of MCG and dust undergoes a sudden
collapse, forming a singularity of the space-time. As expected
there is a decrease in mass due to presence of DE in fig.8. In the
presence of interaction neither DM nor DE(MCG) can dominate over
each other,and hence we observe a mean behaviour between the two
in fig.9. From fig.10 it is evident that there is an increase in
mass with time. The reason being that interaction between DE and
DM alleviates the DE accretion phenomenon on the collapsing
object, thus preventing the mass from decreasing with time. With
GCCG as DE we get almost identical scenario as MCG, in the absence
of interaction with DM, which is evident from figs.11 and 12. In
the presence of interaction, we get a slightly different situation
in case of GCCG as far as mass is concerned. In fig.14 we see that
there is a gradual decrease of mass, thus showing the dominance of
DE over DM, unlike the case of MCG. This is characteristic of
GCCG. The above results are obtained by considering large values
for $r_{c}$, the cross over scale, since for small values of
$r_{c}$, there is no theoretical possibility for a collapse.

~~~~~~~~~~~~~~~~~~~~~~~~~~~~~~~~~~~~~~~~~~~~~~~~~~~~~~~~~~~~~~~~~~~~
~~~~~~~~~~~~~~~~~~~~~~~~~~~~~~~~~~~~~~~~~~~~~~~~~~~~~~~~~~~~~~~~~~~~~~~~~~~~~~~
~~~~~~~~~~~~~~~~~~~~~~~~~~~~~~~~~~~~~~~~~~~~~~~~~~~~~~~~~~~~~~~~~~~~~~~~~~~~~~~~~

~~~~~~~~~~~~~~~~~~~~~~~~~\textbf{LOOP QUANTUM COSMOLOGY
MODEL}~~~~~~~~~~~~~~~~~~~~~

From fig.15 we see that there is at first a gradual decrease and
then a steep decrease of mass due to collapse of DM in LQC model.
This result is totally different from the corresponding result for
DGP brane model. This is indeed quite surprising. The
gravitational pull in case of LQC can be speculated to be really
weak compared to other gravity theories. Fig.16 is quite obvious.
Collapse with MCG in LQC is exhibited in figs. 17 and 18. The
results obtained tally with the results obtained for DGP brane
model. The only difference being that in case of LQC the graph for
$\dot{R}$ rises more steeply than in case of DGP brane. This again
reflects on the weakness of gravity, speculated in the previous
figs. Figs.19 and 20 characterizes the collapse of GCCG in LQC.
Here also the results are similar to that of DGP brane model, the
only difference being the steepness of the slope in LQC just like
the previous graphs. Figs.21 and 22 shows the plots for
combination of MCG and DM in the absence of interaction. Unlike
DGP brane model in fig.21 there is no sudden collapse. In fact the
increase in value of $\dot{R}$ shows the dominating nature of DE
over DM. The large negative pressure of MCG prevails over the weak
gravity of DM, and hence there is a comparatively lesser tendency
of collapse. Fig.22 shows decrease of mass with time, due to MCG
accretion on the collapsing cloud. In presence of interaction MCG
and DM show much moderate behaviour as is evident from the figs.23
and 24. Association with each other suppress their individual
behaviour upto a great extent, thus bringing their mean properties
to light. From figs.25 and 26 we get the characteristics of
collapse with DM and GCCG as DE. Here also unlike DGP brane model
there is no possibility of a sudden collapse, thus showing the
dominance of DE over DM. Finally in figs.27 and 28 we consider
collapse with GCCG and DM in presence of interaction. Unlike the
case of MCG, in this case there is a greater possibility of
collapse as seen from the figs.

Hence in case of LQC, the effect of MCG is felt upto a much
greater extent than the effect of GCCG when combined with DM. This
is characteristic of the two fluids and solely depend on their
respective pressures. The above results are obtained on
considering large values for $\rho_{1}$, the critical loop quantum
density, since for small values of $\rho_{1}$ there is no
theoretical possibility of collapse.

\section{Discussions and Concluding Remarks}\label{chap05}
Here we have studied the gravitational collapse of a spherically
symmetric dust cloud of finite radius, filled with homogeneous and
isotropic fluid. The study was carried out separately in two
different types of modified gravity theories, namely DGP brane
model and Loop quantum cosmology model in the background of a
unified model of DE and DM. Two candidates for DE has been
considered separately, namely modified Chaplygin gas and
Generalized Cosmic Chaplygin gas. At first the effect of DM and
DE(both MCG and GCCG) was studied separately and then their
combined effect was studied both in the presence and absence of
interaction. Values for relevant parameters(time derivative of
radius and mass of the collapsing cloud) were found out and their
variations with time was plotted. A detailed graphical analysis
was done to get a comparison between the results obtained in the
two different models.

It was seen that in case of DM, there is a high possibility for
the matter cloud to undergo collapse, and the consequent formation
of a BH was inevitable, for both DGP and LQC models. In the
background of DE, i.e., MCG or GCCG, the cloud shows a reluctance
to undergo collapse, due to the high negative pressure of the DE
fluids. If at all there is any possibility of a collapse and
subsequent formation of a BH, it will be much delayed in future.
Hence the formation of BH is absolutely an uncertain phenomenon in
this case. This phenomenon is much more pronounced in LQC model
than in the DGP brane model, thus showing the relative weakness of
gravity in the LQC model compared to the DGP brane model. This is
a very interesting result indeed. In the unified model of DE and
MCG in the absence of interaction, there is a sudden collapse
showing the the dominance of matter over DE in the DGP model.
Hence the fate of the collapse is a BH. But we have a completely
different scenario for LQC model, where there is no such collapse
showing the dominance of DE and the weakness of gravity. In the
presence of interaction we see a moderation as far as the
individual effects of DM and DE are concerned. The presence of
either of them suppress the extremeness of their behaviour, as
individuals. In all the above cases the presence of DE decreases
the mass of the cloud due to DE accretion on it. In case of DM
only there is an increase in mass due to accumulation of more
matter.\\\\

{\bf Acknowledgement:}\\
\\
The authors are thankful to IUCAA, Pune, India for warm
hospitality where part of the work was carried out. \\

\end{document}